\documentclass[10pt,journal,compsoc]{IEEEtran}

\newcommand{\toolname}{APPT}

\usepackage{enumerate}
\usepackage{ragged2e}

\usepackage{stfloats}
\usepackage{cite}
\usepackage{amsmath,amssymb,amsfonts}
\usepackage{algorithmic}
\usepackage{graphicx}
\usepackage{textcomp}
\usepackage{xcolor}

\usepackage{url}
\usepackage{hyperref}
\hypersetup{
    colorlinks=true,
    linkcolor=blue,
    filecolor=blue,      
    urlcolor=blue,
    citecolor=blue,
}

\usepackage{multirow}
\usepackage[utf8]{inputenc}
\usepackage{amsmath,amssymb,amsfonts}
\usepackage{algorithmic}
\usepackage{algorithm}
\usepackage{graphicx}
\usepackage{textcomp}
\usepackage{color,xcolor}
\usepackage{url}
\usepackage{graphicx}
\usepackage{subfigure}
\usepackage{stmaryrd}
\usepackage{verbatimbox}
\usepackage{enumerate}
\usepackage[shortlabels]{enumitem}
\usepackage{bm}%

\usepackage{caption}

\usepackage{makecell}
\usepackage{booktabs}

\usepackage{multirow}

\usepackage{enumitem}
\usepackage{colortbl}
\definecolor{codegreen}{rgb}{0,0.6,0}
\definecolor{codegray}{rgb}{0.5,0.5,0.5}
\definecolor{codepurple}{rgb}{0.58,0,0.82}
\definecolor{backcolour}{rgb}{0.95,0.95,0.92}

\usepackage{listings}
\lstdefinestyle{mystyle}{
  backgroundcolor=\color{backcolour},   commentstyle=\color{codegreen},
  keywordstyle=\color{magenta},
  numberstyle=\tiny\color{codegray},
  stringstyle=\color{codepurple},
  basicstyle=\ttfamily\footnotesize,
  breakatwhitespace=false,
  breaklines=true,
  captionpos=b,
  keepspaces=true,
  numbers=left,
  numbersep=5pt,
  showspaces=false,
  showstringspaces=false,
  showtabs=false,
  tabsize=2
}
\newcommand{\finding}[2]{
\begin{center}
\fcolorbox{black}{gray!10}{\parbox{.97\linewidth}{
\textbf{Answer to RQ{#1}:}
{#2}
}}
\end{center}
}

\usepackage[normalem]{ulem}
\newcommand{\revise}[1]{{\color{black}{#1}}}
\newcommand{\delete}[1]{}

\newcommand{\newrevise}[1]{{\color{black}{#1}}}
\newcommand{\newdelete}[1]{}

\AtBeginDocument{%
  \providecommand\BibTeX{{%
    \normalfont B\kern-0.5em{\scshape i\kern-0.25em b}\kern-0.8em\TeX}}}
    
\begin{document}

\title{APPT: Boosting Automated Patch Correctness Prediction \newdelete{using a Pre-trained Language Model}\newrevise{via Fine-tuning Pre-trained Models}}

\author{Quanjun Zhang, Chunrong Fang, Weisong Sun, Yan Liu, Tieke He, Xiaodong Hao, Zhenyu Chen

\IEEEcompsocitemizethanks{
\IEEEcompsocthanksitem 
Quanjun Zhang, Chunrong Fang,  Weisong Sun, Yan Liu, Tieke He, Xiaodong Hao and Zhenyu Chen
are with the State Key Laboratory for Novel Software Technology, Nanjing University, China. \protect\\
E-mail: 
quanjun.zhang@smail.nju.edu.cn,
fangchunrong@nju.edu.cn,
weisongsun@smail.nju.edu.cn,
MF21320104@smail.nju.edu.cn,
hetieke@nju.edu.cn,
MF21320054@smail.nju.edu.cn,
zychen@nju.edu.cn

\IEEEcompsocthanksitem Chunrong Fang, Tieke He, and Zhenyu Chen are the corresponding authors.

}
\thanks{Manuscript received xxx xxx, 2023; revised xxx xxx, 2023.}
}

\markboth{IEEE Transactions on Software Engineering,~Vol.~xxx, No.~xxx, xxx~2024}%
{Shell \MakeLowercase{\textit{et al.}}: Bare Demo of IEEEtran.cls for Computer Society Journals}

\IEEEtitleabstractindextext{
\begin{abstract}
\justifying
Automated program repair (APR) aims to fix software bugs automatically without human debugging efforts and plays a crucial role in software development and maintenance.
Despite the recent significant progress in the number of fixed bugs, APR is still challenged by a long-standing overfitting problem (i.e., the generated patch is plausible but overfitting).
Various techniques have thus been proposed to address the overfitting problem.
\newdelete{Among them, leveraging deep learning approaches to predict patch correctness automatically is emerging along with the available large-scale patch benchmarks recently.
However, existing learning-based techniques usually adopt a two-fold prediction pipeline, i.e., the feature extractor and the classifier.
The two components are isolated and only the classifier is trained, which may limit the prediction effectiveness.}
\newrevise{Recently, researchers have employed BERT to extract code features, which are then used to train a classifier for patch correctness prediction, indicating the potential of such pre-trained models in reasoning about patch correctness.
However, BERT is restricted to feature extraction for classifier training without benefiting from the training process, potentially generating sub-optimal vector representations for patched code snippets.}
\newdelete{In this paper, we propose {\toolname}, a pre-trained model-based automated patch correctness assessment technique, which adopts a pre-trained model as the encoder stack, followed by an LSTM stack and a deep learning classifier.
More importantly, the encoder stack is fine-tuned with other components as a whole pipeline to fully adapt the pre-training model for reasoning about the patch correctness domain.}
\newrevise{In this paper, we propose {\toolname}, a pre-trained model-based automated patch correctness assessment technique by both pre-training and fine-tuning.
APPT adopts a pre-trained model as the encoder stack, followed by an LSTM stack and a deep learning classifier.
More importantly, the pre-trained model is fine-tuned in conjunction with other components as a whole pipeline to fully adapt it specifically for reasoning about patch correctness.}
Although our idea is general and can be built on various existing pre-trained models, we have implemented {\toolname} based on the BERT model.
We conduct an extensive experiment on 1,183 Defects4J patches and the experimental results show that {\toolname} achieves prediction accuracy of 79.7\% and recall of 83.2\%, outperforming the state-of-the-art technique CACHE by 4.3\% and 6.7\%.
Our additional investigation on 49,694 real-world patches shows that {\toolname} achieves the optimum performance (exceeding 99\% in five common metrics for assessing patch classification techniques) compared with existing representation learning techniques.
\revise{We further investigate the impact of each component and find that they all positively contribute to {\toolname}, e.g., the fine-tuning process and the LSTM stack increase F1-score by 10.22\% and 4.11\%, respectively.}
We also prove that adopting advanced pre-trained models can further provide substantial advancement (e.g., GraphCodeBERT-based {\toolname} improves BERT-based {\toolname} by 2.8\% and 3.3\% in precision and AUC, respectively), highlighting the generalizability of {\toolname}.
\newrevise{Overall, our study highlights the promising future of fine-tuning pre-trained models 
to assess patch correctness and reduce the manual inspection effort of debugging experts when deploying APR tools in practice.}

\end{abstract}

}

\maketitle
\IEEEdisplaynontitleabstractindextext

\IEEEpeerreviewmaketitle

\IEEEraisesectionheading{
\section{Introduction}
\label{sec:intro}
}
\IEEEPARstart{S}oftware bugs are inevitable in modern software systems and result in fatal consequences, 
such as costing trillions of dollars in financial loss and affecting billions of people around the world~\cite{gazzola2017automatic,benton2021evaluating}.
It is incredibly time-consuming and labor-intensive for developers to fix such bugs due to the increasing size and complexity of modern software systems~\cite{2009Aranda,winter2022let}.
Automated program repair (APR) aims to fix revealed software bugs without human intervention automatically and has attracted massive attention from both academia and industry in the past decades~\cite{goues2019automated,kirbas2021introduction,winter2022towards}.
Despite an emerging research area, a variety of APR techniques have been proposed and continuously achieved promising results in terms of the number of fixed bugs in the literature~\cite{durieux2019empirical,liu2020efficiency}.

However, it is fundamentally difficult to achieve high precision for generated patches due to the weak program specifications \cite{le2019reliability}.
Existing APR techniques usually leverage the developer-written test cases as the criteria to assess the correctness of the generated patches.
In fact, a generated patch passing the available test cases may not generalize to other potential test cases, leading to a long-standing challenge of APR (i.e., the overfitting issue) \cite{zhang2023survey,le2019reliability}.
For example, when a bug is detected in functionality, a patch can be simply generated by deleting the functionality and the available test cases usually fail to exercise the deleted functionality \cite{long2016analysis}.
In this case, developers need to consume tremendous time and effort to filter the overfitting patches, resulting in a negative debugging performance when APR techniques are applied in practice \cite{tao2014automatically, zhang2022program,zhang2022interactive}.

Thus, various automated patch correctness assessment (APCA) techniques have been proposed to determine whether a generated patch is indeed correct or not \cite{tian2020evaluating}.
According to extracted features, the traditional APCA techniques can be categorized into two groups: static and dynamic ones\cite{wang2020automated}.
Static techniques tend to analyze the code changed patterns or code similarity based on the syntactic and semantic features.
For example, Tan et al. \cite{tan2016anti} define a set of generic forbidden transformations (e.g., the above-mentioned functionality deleting) for the buggy program.
In contrast, dynamic techniques usually execute the plausible patches against extra test cases generated by automated test generation tools (e.g., Evosuite \cite{fraser2011evosuite} and Randoop \cite{pacheco2007randoop}).
For example, Xiong et al. \cite{xiong2018identifying} generate new test cases and determine patch correctness based on the behavior similarity of the test case executions.
However, the static techniques may suffer from prediction precision problems, while it is pretty time-consuming for dynamic techniques to generate additional test cases and execute all patched programs~\cite{wang2020automated}.

\newdelete{Recently, inspired by large-scale patch benchmarks being released \cite{liu2020efficiency,durieux2019empirical}, some learning-based APCA techniques have been proposed to assess patch correctness by embedding buggy and patched code snippets \cite{ye2021automated,tian2020evaluating,tian2022predicting}.
For example, He et al. \cite{ye2021automated} extract hand-crafted features from Java programs statically and train a probabilistic model (i.e., ODS) to perform patch correctness prediction.
Despite appealing, the construction of carefully hand-crafted  features requires professional knowledge in the domain and it is challenging to generalize such features to other scenarios (e.g., different programming languages) \cite{tian2022best}.
Instead of hand-crafted features,  Tian et al. \cite{tian2020evaluating} investigate the feasibility of embedding features to build predictive models.
However, despite outstanding prediction results, Tian et al. \cite{tian2020evaluating} share a similar two-fold prediction workflow to ODS, i.e., the feature extractor and the classifier. 
The two components are independent and the feature extractor is only responsible for extracting features from code snippets. 
As a result, only the parameters of classifiers are optimized during training, limiting the prediction effectiveness.
}

\newrevise{
Recently, inspired by large-scale patch benchmarks being released \cite{liu2020efficiency,durieux2019empirical}, some learning-based APCA techniques have been proposed to assess patch correctness by embedding buggy and patched code snippets \cite{ye2021automated,tian2020evaluating,tian2022predicting}.
For example, He et al. \cite{ye2021automated} extract hand-crafted features from Java programs statically and train a probabilistic model (i.e., ODS) to perform patch correctness prediction.
Despite appealing, the construction of carefully hand-crafted features requires professional knowledge in the domain and it is challenging to generalize such features to other scenarios (e.g., different programming languages) \cite{tian2022best}.
Instead of hand-crafted features,  Tian et al. \cite{tian2020evaluating} investigate the feasibility of embedding features to build predictive models.
Among investigated embedding models, the pre-trained BERT achieves optimal results, demonstrating the potential of such pre-trained models in reasoning about patch correctness.
However, BERT is only responsible for extracting features to train the classifier and does not benefit from the training process.
Besides, such pre-trained models are usually trained to derive generic knowledge by self-supervised training on various types of corpora (e.g., natural language), thus may generate sub-optimal vector representations for patched code snippets, limiting prediction performance.}

\textbf{This Paper.}
In this work, we propose, \textbf{\textit{{\toolname}}}, \delete{the first }\revise{an} \underline{\textbf{\textit{A}}}utomated \underline{\textbf{\textit{P}}}re-trained model-based \underline{\textbf{\textit{P}}}atch correc\underline{\textbf{\textit{T}}}ness assessment technique, which employs \revise{both} the pre-training and fine-tuning to address the above limitation of prior work.
We first adopt the large pre-trained model as the encoder stack to extract code representations.
We then employ bidirectional LSTM layers to capture rich dependency information between the buggy and patched code snippets.
Finally, we build a deep learning classifier to predict whether the patch is overfitting or not.
Unlike ODS, {\toolname} treats only the source code tokens as the input and automatically extracts code features using a well-trained encoder stack, getting rid of the need for manually-designed features \cite{ye2021automated}.
Besides, different from Tian et al. \cite{tian2020evaluating} only training the classifiers, APPT is able to further fine-tune the BERT model to obtain the optimal embedding vectors for patch correctness.
\newrevise{This domain adaptation is expected to make the model's representations more relevant to distinguishing correct from overfitting patches.}
Although {\toolname} is conceptually general and can be built on various pre-trained models, we have implemented {\toolname} as a practical APCA tool based on the BERT model.
Our experimental results on 1,183 Defects4J patches indicate that {\toolname} improves the state-of-the-art technique CACHE by 4.36\% accuracy, 1.3\% precision, 6.7\% recall, 3.8\% F1-score and 2.2\% AUC.
We conduct an additional investigation on 49,694 real-world patches from five different patch benchmarks and the results show that {\toolname} exceeds 99\% in accuracy, precision, recall, F1-score and AUC metrics, outperforming the existing representation learning techniques.
\revise{
Our ablation study demonstrates that the components in {\toolname} all positively contribute to {\toolname}.
For example, the improvement achieved by the fine-tuning process reaches 1.82\%$\sim$13.16\% for all metrics.}
We adopt different pre-trained models to further investigate the generalization ability of {\toolname}.
The results demonstrate that {\toolname} with advanced pre-trained models can enhance the prediction performance.
For example, precision and AUC of {\toolname} can be improved by 2.8\% and 3.3\% when equipped with GraphCodeBERT, which are 4.2\% and 5.4\% higher than the state-of-the-art technique CACHE.

To sum up, we make the following major contributions:
\begin{itemize}
  \item \newdelete{\textbf{New Direction.}
  This paper opens a new direction for patch correctness assessment to \delete{directly }utilize large pre-trained models by \revise{both} pre-training and fine-tuning. 
  \delete{Compared with existing learning-based APCA techniques, our approach does not need any additional efforts to design and extract complex code features.}
  Compared with existing representation learning techniques, the pre-trained model is further fine-tuned with other components in {\toolname} architecture to obtain optimal embedding vectors for reasoning \newrevise{about} patch correctness.}
  
  \newrevise{\textbf{Prediction Pipeline.}
  This paper introduces a prediction pipeline for patch correctness assessment, leveraging large pre-trained models through a process of pre-training followed by fine-tuning.
  Compared with existing representation learning techniques, the pre-trained model is further fine-tuned with other components in {\toolname} architecture to obtain optimal embedding vectors for reasoning about patch correctness.}

  \item \textbf{Novel Technique.}
  We propose {\toolname}, a BERT-based APCA technique that leverages the pre-training and classifier to predict patch correctness.
  To the best of our knowledge, we are the first to exploit fine-tuning the pre-trained model for assessing patch correctness.
  
  \item \textbf{Extensive Study.}
  We conduct various empirical studies to investigate and evaluate {\toolname} on diverse patch benchmarks. The results show that {\toolname} achieves significantly better overall performance than existing learning-based and traditional APCA techniques.
  
  \item \textbf{Available Artifacts.}
  We release the relevant materials (including source code, patches and results) used in the experiments for replication and future research\footnote{All artifacts relevant to this work can be found at \url{https://github.com/iSEngLab/APPT}, accessed March 2023.}.

\end{itemize}

\section{Background}
\label{sec:bg&mv}
\subsection{Automated Program Repair}

\begin{figure}[t]
\centering
\graphicspath{{graphs/}}
    \includegraphics[width=0.95\linewidth]{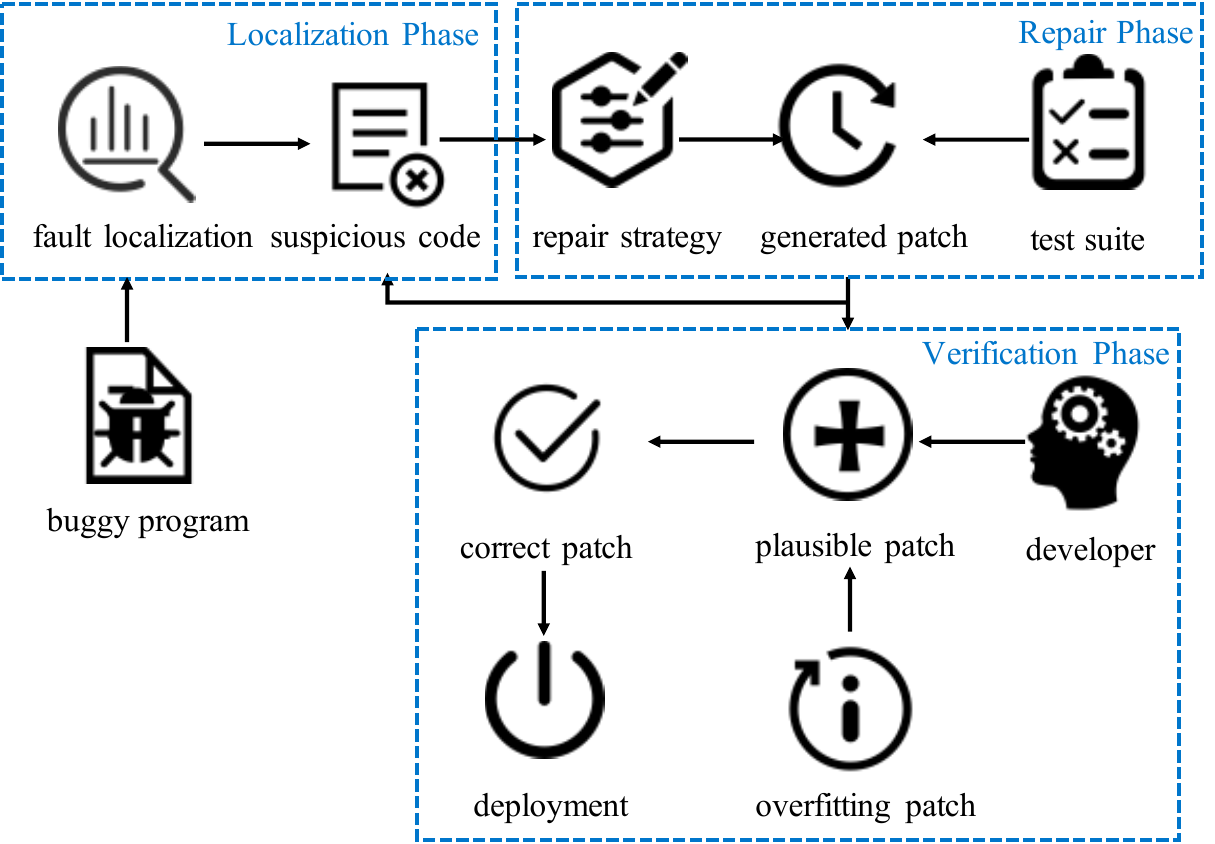}
    \caption{Overview of APR}
    \label{fig:apr}
\end{figure}

APR techniques' primary objective is to identify and fix program bugs automatically.
Fig. \ref{fig:apr} illustrates the workflow of the typical APR technique, which is usually composed of three steps:
(1) the localization \newdelete{phrase}\newrevise{phase} utilizes off-the-shelf fault localization techniques to recognize the suspicious code elements (e.g., statements or methods) \cite{2016Wong,lou2020can};
(2) the repair \newdelete{phrase}\newrevise{phase} then modifies these elements based on a set of transformation rules to generate various new program variants, also called candidate patches;
(3) the verification \newdelete{phrase}\newrevise{phase} adopts the original test cases as the oracle to check whether candidate patches execute as expected or not.
Specifically, a candidate patch passing the original test cases is called a \emph{plausible} patch.
A plausible patch that is semantically equivalent to the developer patch denotes a \emph{correct} patch; otherwise, it is an \emph{overfitting} patch.

It is fundamentally challenging to ensure the correctness of the plausible patches due to the weak specification of the program behavior in practice.
Existing studies have demonstrated that manually identifying the overfitting patches is time-consuming and may harm the debugging performance of developers \cite{tao2014automatically, smith2015cure}.
Thus, various techniques have been proposed to validate patch correctness automatically.
According to whether the dynamic execution or machine learning is required \cite{wang2020automated}, we categorize them into three main categories:
static-based techniques, dynamic-based techniques and learning-based techniques.

\emph{$\bullet$ Static-based APCA techniques.}
These techniques aim to prioritize correct patches over overfitting ones by static code features, such as code-deleting program transformations.

\emph{$\bullet$ Dynamic-based APCA techniques.}
These techniques aim to filter out overfitting patches by executing extra test cases, which are generated based on fixed or patched programs.
According to whether the correct patches are required, these techniques can be further categorized into \emph{dynamic with oracle-based ones} and \emph{dynamic without oracle-based ones}.

\emph{$\bullet$ Learning-based APCA techniques.}
These techniques aim to predict the correctness of plausible patches enhanced by machine learning techniques.
They usually extract the manually-designed code features and then adopt a classifier to perform patch prediction \cite{ye2021automated}.
Some techniques are proposed to adopt code embedding techniques to extract code features automatically \cite{lin2021context}, which are also denoted as \emph{representation learning-based APCA techniques}.

Recently, an increasing number of research efforts have attempted to use machine learning techniques to learn from existing patch benchmarks for predicting potential patch correctness, achieving promising results.
In this work, we adopt the large pre-trained model (i.e., BERT) to encode plausible patches and train a deep learning classifier to predict patch correctness.
Compared to existing techniques, our paper is the first work to predict patch correctness by pre-training and fine-tuning the pre-trained model.

\subsection{Pre-trained Model}
Recently, Pre-trained\newdelete{ language} models (e.g., BERT) have significantly improved performance across a wide range of natural language processing (NLP) tasks, such as machine translation and text classification \cite{devlin2018bert,brown2020gpt,raffel2019t5}.
Typically, the models are pre-trained to derive generic language representations by self-supervised training on large-scale unlabeled data and then are transferred to benefit multiple downstream tasks by fine-tuning on limited data annotation.

Existing pre-trained models usually adopt the encoder-decoder architectures, where an encoder encodes an input sequence as a fixed-length vector representation, and a decoder generates an output sequence based on the input representation.
Encoder-only models (e.g., BERT \cite{devlin2018bert}) usually pre-train a bidirectional transformer in which each token can attend to each other.
Encoder-only models are good at understanding tasks (e.g., code search), but their bidirectionality nature requires an additional decoder for generation tasks, where this decoder initializes from scratch and cannot benefit from the pre-training tasks.
Decoder-only models (e.g., GPT \cite{brown2020gpt}) are pre-trained using unidirectional language modeling that only allows tokens to attend to the previous tokens and itself to predict the next token.
Decoder-only models are good at auto-regressive tasks like code completion, but the unidirectional framework is sub-optimal for understanding tasks.
Encoder-decoder models (e.g., T5 \cite{raffel2019t5}) often make use of denoising pre-training objectives that corrupt the source input and require the decoder to recover them.
Compared to encoder-only and decoder-only models that favor understanding and auto-regressive tasks, encoder-decoder models can support generation tasks like code summarization. 
In this work, we treat the patch correctness assessment as a binary classification task, and we consider encoder-only models to get embeddings of code snippets according to existing work \cite{guo2022unixcoder}.
\newrevise{Our focus is to investigate the potential of transferring the rich general knowledge acquired during the pre-training phase to the downstream task (i.e., reasoning about patch correctness) via fine-tuning.}

Inspired by the success of pre-trained models in NLP, many recent attempts have been adopted to boost numerous code-related tasks (e.g.,  code summarization and code search) with pre-trained models (e.g., GraphCodeBERT) \cite{zhang2023survey2}.
Despite the promising results, little work aims to explore the capabilities of pre-trained models in supporting patch correctness assessment.
In this work, BERT is selected to exploit pre-trained models for automated patch correctness assessment, as it has been widely adopted in various code-related tasks and is quite effective for classification tasks \cite{feng2020codebert, guo2020graphcodebert}.
Two advanced BERT-style models (i.e., CodeBERT and GraphCodeBERT) are also selected to investigate the generalization ability of {\toolname}.

\section{Approach}
\label{sec:approach}
\begin{figure*}[htbp]
\centering
\graphicspath{{graphs/}}
    \includegraphics[width=0.99\linewidth]{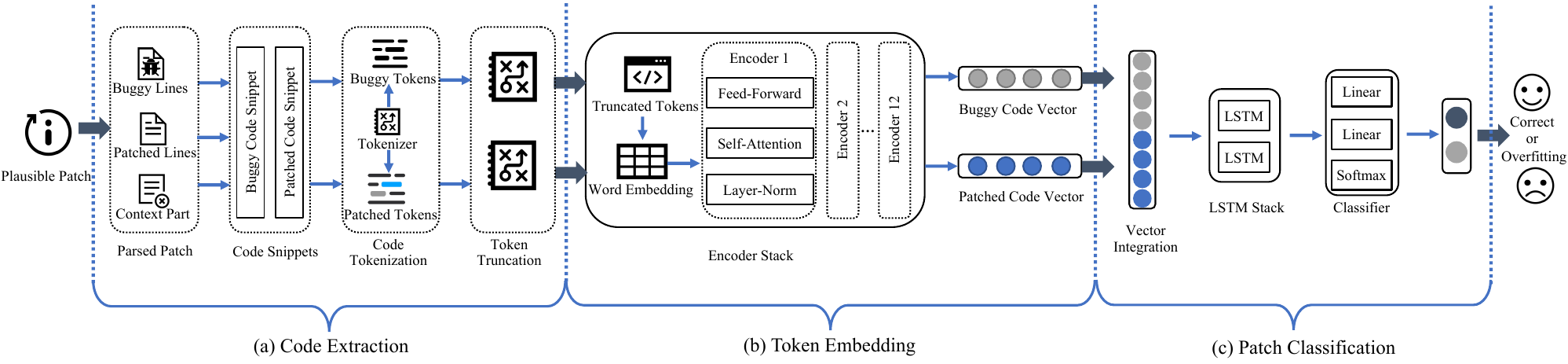}
    \caption{Overview of {\toolname}}
    \label{fig:framework}
\end{figure*}

Fig. \ref{fig:framework} presents the overall framework of our approach. 
Generally, {\toolname} accepts a buggy program and a plausible patch that passes the available test cases as inputs.
\delete{
{\toolname} extracts the buggy code snippet and its corresponding patched code snippet, and adopts four strategies to truncate the code tokens.}
\revise{{\toolname} extracts the buggy code snippet and its corresponding patched code snippet, and truncates the code tokens for embedding.}
{\toolname} then uses the pre-trained BERT model for embedding the truncated tokens.
After obtaining the representations for the buggy and patched code snippets, {\toolname} uses four pre-defined functions for integrating the representations.
Finally, {\toolname} adopts a deep learning classifier to return the final result (i.e., correct or overfitting).

\subsection{Code Extraction}
Given a buggy program, existing APR tools may return a plausible patch $p$ (if it exists) that passes all available test cases.
\textit{Code extraction \newdelete{phrase}\newrevise{phase}} aims to take the returned patch and the buggy program as the inputs, and output the corresponding buggy and patched code tokens (shown in Fig. \ref{fig:framework}(a)).

Specifically, we get the buggy and patched code snippets (i.e., $C_b$ and $C_p$) by  parsing the patch file.
Firstly, we select removed and added lines as the buggy and patched lines, marked with ``{\color{green}{+}}'' and \delete{‘}\revise{``}{\color{red}{-}}\delete{’}\revise{''}, respectively.
Secondly, to keep the context information about the plausible patch, we keep unchanged lines (i.e., without \revise{``}{\color{green}{+}}'' and \delete{‘}\revise{``}{\color{red}{-}}\revise{''} in the beginning) as part of each code snippet.
Finally, the buggy (or patched) code snippet \delete{are}\revise{is} made up \delete{by}\revise{of} the buggy (patched) lines and common context part.

We treat the buggy (or patched) code snippet as sequences of tokens and utilize a subword tokenization method to address the out-of-vocabulary problem by breaking down identifiers into their subtokens \cite{jiang2021cure} when tokenizing the code snippet.
In this work, we keep the original tokenization vocabulary instead of building a new vocabulary using the byte-pair-encoding algorithm, such that {\toolname} can inherit the natural language understanding ability and start learning prediction from a good initial point.

After the buggy (or patched) code tokens are extracted, we attempt to take them as inputs into the token embedding \newdelete{phrase}\newrevise{phase}.
However, pre-trained models are usually limited to a particular input length.
For example, BERT can only take input sequences up to 512 tokens in length.
We further truncate the inputs whose length is longer than 512 after tokenization.
\newdelete{In such a case, we directly keep the first 512 tokens in $C_b$ and $C_p$.
Finally, the buggy and patched code tokens (i.e., $T_b$ and $T_p$) are extracted based on $C_b$ and $T_p$ to fit the maximum length limit of BERT.}
\newrevise{In particular, we first parse the given buggy code snippet $C_b$ (and the patched one $T_p$) into individual code tokens $T_b$ (and $T_p$).
We then check whether the length of the code snippet $C_b$ (and $T_p$) exceeds 512 tokens and retain only the first 512 tokens in $T_b$ ( and $T_p$) to represent the code snippet.
Finally, the buggy and patched code tokens (i.e., $T_b$ and $T_p$) are extracted and truncated based on $C_b$ and $T_p$ to fit the maximum length limit of BERT.
}
\delete{
Following existing work \cite{ganhotra2021does}, we use different methods to truncate the method pair.}

\delete{
head-only: keep the first 512 tokens in $C_b$ and $C_p$.}

\delete{tail-only: keep the last 512 tokens in $C_b$ and $C_p$.}

\delete{mid-only: select 512 tokens in the middle of in $C_b$ and $C_p$.}

\delete{hybrid: select the first 256 and the last 256 tokens in $C_b$ and $C_p$.}

\delete{
In our experiment, we use the head-only method to truncate the code tokens by default.
We also discuss the impact of different truncation methods in Section \ref{sec:rq3.1}.
Finally, the buggy and patched code tokens (i.e., $T_b$ and $T_p$) are extracted based on $C_b$ and $T_p$ to fit the maximum length limit of BERT.
}

\subsection{Token Embedding}
\textit{Token Embedding \newdelete{phrase}\newrevise{phase}} takes the buggy (or patched) code tokens (i.e., $T_b$ or $T_p$) as input and embeds it into the buggy (or patched) vector (i.e., $E_b$ or $E_p$) as output (shown in Fig. \ref{fig:framework}(b)).
{\toolname} implements a stack of twelve layers of encoder blocks to extract the hidden state of the code snippet.
Each encoder block consists of three components.
The first part is a multi-head self-attention layer to learn long-range dependencies in the input code tokens.
The second part is a simple, position-wise fully connected feed-forward neural network, which can linearly transform the token embedding for better feature extraction. 
The third part is a residual connection around each component, followed by a layer normalization to ensure the stability of code token embeddings distribution.

In particular, the self-attention mechanism computes the representation of each code token by considering the position relationship between the code tokens.
It mainly relies on three main vectors, query $Q$, key $K$, and value $V$, by mapping a query and a set of key-value pairs to an output vector.
We employ a scaled dot-product self-attention to calculate the attention scores of each token by taking the dot product between all of the query vectors and key vectors. 
The attention scores are then normalized to probabilities using the softmax function to get the attention weights.
Finally, the value vectors can be updated by taking a dot product between the value vectors and the attention weight vectors.
The self-attention operation is computed using three matrices $Q$, $K$ and $V$ as follows:

\begin{equation}
\operatorname{Attention}(Q, K, V)=\operatorname{softmax}\left(\frac{QK^{T}}{\sqrt{d_{k}}}\right)V
\end{equation}

To capture richer semantic meanings of the input code tokens, we further use a multi-head mechanism to realize the self-attention, which allows the model to jointly attend the information from different code representation subspaces at different positions.
 For $d$-dimension $Q$, $K$, and $V$, we split those vectors into $h$ heads where each head has $d/h$-dimension. 
 After all of the self-attention operations, each head will then be concatenated back again to feed into a fully-connected feed-forward neural network including two linear transformations with a ReLU activation in between. 
The multi-head mechanism can be summarized by the following equation:

\begin{equation}
\small
\operatorname { MultiHead }(Q, K, V)=\operatorname { Concat }\left(\operatorname{head}_{1}, \ldots, \operatorname{head}_{h}\right) W^{O}
\end{equation}
where $head_i=Attention(QW^Q_i,KW^Q_i,VW^Q_i)$ and $W^{O}$ is used to linearly project to the expected dimension after concatenation.
Therefore, the encoder stack can take an input code snippet and output a real-valued vector for each code token within the code snippet based on the context.
\revise{Besides, to obtain optimal embedding vectors in the APCA domain, the encoder stack is further fine-tuned with other components as a whole prediction pipeline.}

\subsection{Patch Classification}
\label{sec:classification}

After the embedding vectors of the buggy and patched code snippets (i.e., $E_b$ and $E_p$) are extracted by the encoder stack, \textit{patch classification \newdelete{phrase}\newrevise{phase}} first integrates the two vectors into a single input vector (i.e., $E_{con}$) and then adopts a deep learning classifier to predict the patch correctness automatically (shown in Fig. \ref{fig:framework}(c)).

\subsubsection{Representations Integration}
\label{sec:vector}

Given two vectors $E_b$ and $E_p$ with $n$ dimensions representing the buggy and patched code snippets, respectively, we integrate the two vectors into one code-changed vector for patch classification.
\revise{
There exist different approaches to integrate $E_b$ and $E_p$ with the aim to characterize their differences from diverse aspects, such as a vector-wise concatenation operation $E_{con}$, element-wise addition operation $E_{add}$, element-wise subtraction operation $E_{sub}$, Hadamard product $E_{pro}$.
}
\begin{enumerate}[(1)]
  \item \bm{$E_{con}$} is a concatenation operation between $E_b$ and $E_p$ on vector-wise level with $2n$ dimension (i.e., $E_{con} = E_b \bigoplus E_p$). 
  \item \bm{$E_{add}$} is an addition operation between $E_b$ and $E_p$ on element-wise level with $n$ dimensions (i.e., $E_{add} = E_b + E_p$).
  \item \bm{$E_{sub}$} is a subtraction operation between $E_b$ and $E_p$ on element-wise level with $n$ dimensions (i.e., $E_{sub} = E_b - E_p$).
  \item \bm{$E_{pro}$} is a Hadamard product operation between $E_b$ and $E_p$ on element-wise level with $n$ dimensions (i.e., $E_{sub} = E_b \odot E_p$).
\end{enumerate}

\revise{
The concatenation operation is proven to be simple yet effective \newdelete{in previous feature fusion studies \cite{zhang2020multimodal, sleeman2022multimodal}}\newrevise{in previous work~\cite{tian2022best}}.
Besides, it does not lose the original $E_b$ and $E_p$ vector values, which allows the following LSTM stack to further extract deep relationships between them.
Thus, {\toolname} employs the concatenation operation as its default vector integration method.
In principle, {\toolname} can also leverage other approaches (e.g., addition and subtraction) for vector integration.
We would further investigate the impacts of different vector integration approaches in the detailed experiments.
}

\delete{
In detail, we leverage different approaches to integrate them to characterize the differences between $E_b$ and $E_p$ from diverse aspects, such as \delete{an} \revise{a} vector-wise concatenation operation $E_{con}$, element-wise addition operation $E_{add}$, element-wise subtraction operation $E_{sub}$, Hadamard product $E_{pro}$.
We also attempt to capture crossed features between the two vectors by concatenating the \delete{above integrated} \revise{above-integrated} vectors $E_{mix}$.
The integration approaches are selected due to their promising results in previous studies \cite{hoang2020cc2vec, tian2020evaluating}, which are listed as follows:
}

\subsubsection{LSTM Stack}
After the embedding vector (e.g., $E_{con}$) of the changed code tokens is extracted, 
{\toolname} aims to determine the given patch's correctness based on a deep learning classifier.
To extract more hidden code change features, we further feed the code changed vector into a Long Short-Term Memory (LSTM) stack.
The LSTM stack has two bidirectional LSTM layers, the output of which is a new state generated by concatenating the hidden states from both directions at a time.
LSTM is a specialized recurrent neural network (RNN) for modeling long-term dependencies of sequences.
A common LSTM gate unit is composed of a cell, an input gate, an output gate and a forget gate.
Thanks to the gated mechanism, LSTM is well-suited to extract the contextual semantic features containing token sequential dependencies and has been widely used in various kinds of tasks, such as vulnerability detection \cite{li2018vuldeepecker}, fault localization \cite{meng2022improving}, and automated program repair \cite{li2020dlfix}.
\newrevise{Considering that patches may contain changes that are non-adjacent but related, the LSTM stack is adept at capturing these long-range dependencies within the data, which could be instrumental in assessing the correctness of a patch.}

In {\toolname}, the LSTM stack computes a mapping from an input code changed vector $x = (x_1, ..., x_T)$ (e.g., $E_{con}$) to an output vector $z = (z_1, ..., z_T)$ by calculating the network gate unit activations.
We implement the gated mechanism by leveraging the input gates and forget gates to control the propagation of cell states.
Specifically, when updating the cell state, the input gates decide what new information from the current input to be included in the cell states (i.e., Equation \ref{equ:ig}), and forget gates decide what information to be excluded from the cell states (i.e., Equation \ref{equ:fg}).
Based on new and forgetting information, cell states as the memory of the LSTM unit can be updated (i.e., Equation \ref{equ:cs}).
The output gate then determines the value for the next hidden state by point-wise multiplication of the output gate (i.e., Equation \ref{equ:og}).
Finally, the value of the current cell state passed through tanh function (i.e., Equation \ref{equ:hs}), by which the output of LSTM stack is calculated (i.e., Equation \ref{equ:out}). 

\begin{equation}
\label{equ:ig}
i_{t}=\operatorname{sigmoid}\left(W_{ix} x_{t} + W_{ih} h_{t-1} + b_{i}\right) \\
\end{equation}

\begin{equation}
\label{equ:fg}
f_{t}=\operatorname{sigmoid}\left(W_{fx} x_{t}+W_{fh} h_{t-1}+b_{f}\right) \\
\end{equation}

\begin{equation}
\label{equ:cs}
c_{t}=f_{t} \odot c_{t-1}+i_{t} \odot \tanh \left(W_{gx} x_{t}+W_{gh} h_{t-1}+b_{g}\right) \\
\end{equation}

\begin{equation}
\label{equ:og}
o_{t}=\operatorname{sigmoid}\left(W_{ox} x_{t}+W_{oh} h_{t-1}+b_{o}\right) \\
\end{equation}

\begin{equation}
\label{equ:hs}
h_{t}=o_{t} \odot \tanh \left(c_{t}\right)
\end{equation}

\begin{equation}
\label{equ:out}
z_{t}=W_{zh} h_{t}+b_{z} \\
\end{equation}
where the $W$ terms denote weight matrices (e.g., $W_{ix}$ is the matrix of weights from the input gate to the input), the $b$ terms denote bias vectors (e.g., $b_i$ is the input gate bias vector) and $\odot$ denotes element-wise multiplication of the vectors.

\subsubsection{Classifier}
After the computation of all LSTM iterations, the embedding vectors of changed code tokens are further fed to a designed deep learning classifier to predict the patch correctness.
The classifier is composed of two fully connected layers followed by a binary predictor. 
In {\toolname}, we apply a standard softmax function to obtain the probability distribution over correctness. 
A patch is labeled as correct if its probability of being correct is larger than that of being incorrect; otherwise, it is considered overfitting.

In particular, for patch $p$, $z$ denotes its output of the last iteration in the LSTM stack, which is further linearly transformed into a real number as Equation \ref{equ:liner}, where $W \in \mathbb{R}^{d \times 1}$, $b \in \mathbb{R}$, and $n$ denotes the number of class (i.e., correct and overfitting).
We then leverage softmax function to normalize the output of patch $p$ as Equation \ref{equ:normalize}, where $s$ denotes the correct or overfitting probability of patch $p$ predicted by the model.
\revise{A patch is considered correct if its probability of being correct is larger than that of being incorrect (i.e., larger than 0.5); otherwise, it is considered overfitting.}

\begin{equation}
\label{equ:liner}
y_i=W z_i + b_i 
\quad \forall \mathrm{i} \in 1 \ldots  \mathrm{n}
\end{equation}

\begin{equation}
\label{equ:normalize}
s\left(y_i\right)=\frac{\exp \left\{y_{i}\right\}}{\sum_{i=1}^{n} \exp \left\{y_{j}\right\}}
\end{equation}

\subsection{Training}

To train the network, we calculate the loss to update the neural weights based on its predicted result and ground truth.
We use the cross-entropy loss, which has been widely used in some classification tasks and patch prediction studies \cite{lin2021context, alon2019code2vec}. 
In particular, $g_i \in \{0,1\}$ denotes whether the $i$-th patch is correct or overfitting.
The cross-entropy loss compares a target $g_i$ with a prediction $s$ in a logarithmic and hence exponential fashion.
The objective function is computed in Equation \ref{equation:loss}, which is minimized constantly in the training to update the parameters in our model.
\revise{It should be noted that, different from Tian et al. \cite{tian2020evaluating} only training the classifier, the whole architecture (i.e., the encoder stack, the LSTM stack, and the clasifier) in our APPT is optimized during training.}

\begin{equation}
L=\sum_{i}-[g_{i} \cdot \log (s)+(1-g_{i}) \cdot \log (1-s)]
\label{equation:loss}
\end{equation}

We employ the dropout technique to improve the robustness of {\toolname} and the Adam approach \cite{kingma2014adam} to optimize the objective function.

\section{Experiment}
\label{sec:exp}

\subsection{Research Questions}
\delete{
The empirical study is conducted to answer the following research questions.}
\revise{{\toolname} is designed to predict patch correctness among a mass of plausible patches automatically. 
To this end, we explore the following research questions (RQ):}

\begin{description}
\item	

    \item [\revise{RQ1 (Effectiveness):}] 
    \delete{How does {\toolname} perform compared with existing state-of-the-art representation learning-based APCA techniques?}
    \revise{How does {\toolname} perform compared with existing state-of-the-art APCA techniques?}
    \begin{description}
        \item [\revise{RQ1.1:}]
        \revise{How does {\toolname} perform compared with existing state-of-the-art representation learning-based APCA techniques?}
        \item [\revise{RQ1.2:}]
        \revise{How does {\toolname} perform compared with existing state-of-the-art traditional and learning-based APCA techniques?}
    \end{description}

    \item [RQ2 (Impact analysis):] To what extent do the different choices affect the overall effectiveness of {\toolname}?
    \begin{description}
    \revise{
        \item [RQ2.1:]\delete{To what extent do the token truncation choices affect the overall effectiveness of {\toolname}?}\revise{To what extent do the training choices affect the overall effectiveness of {\toolname}?}}
        \item [RQ2.2:]To what extent do the vector concatenation choices affect the overall effectiveness of {\toolname}?
        \item [RQ2.3:]To what extent do the pre-trained model choices affect the overall effectiveness of {\toolname}?
    \end{description}

    \item [\revise{RQ3 (Cross-project effectiveness):}]
    \revise{How does {\toolname} perform on the new projects in the cross-project prediction scenario?}

\end{description}

\revise{
RQ1 aims to investigate the effectiveness of {\toolname}, which is further refined into two sub-RQs.
In detail, RQ1.1 explores to what extent {\toolname} outperforms existing representation learning techniques, including three classifiers multiplied (decision tree, logistic regression, and naive Bayes) by five representation methods (BERT, code2vec, code2seq, Doc2Vec, and CC2Vec) from Tian et al. \cite{tian2020evaluating}, and the most recent technique CACHE from Lin et al. \cite{lin2021context}.
RQ2.2 explores the effectiveness of {\toolname} by comparing it with both dynamic and static techniques.
The latest learning-based APCA technique, ODS, is also evaluated in our study.
RQ2 focuses on impact analysis of {\toolname}, which is further refined into three sub-RQs.
In detail, RQ2.1 explores how the training choices affect the effectiveness of {\toolname}.
RQ2.2 explores how the five vector concatenation
methods affect the effectiveness of {\toolname}.
RQ2.3 replaces BERT with advanced CodeBERT and GraphCodeBERT to investigate the impact of the pre-trained models on the effectiveness of {\toolname}.
RQ3 explores the effectiveness of {\toolname} when predicting patches from unseen projects.
}

\subsection{Dataset}

\delete{
With the rapid development of APR research in the last decades, a broad range of repair techniques has been proposed \cite{Yuan2018Arja, lutellier2020coconut,yuan2022circle}, resulting in a growing number of patches across many benchmarks being released \cite{liu2020efficiency,wang2020automated}.
The large-scale patch benchmarks enable deep learning-based prediction techniques to learn the distribution of correct and overfitting patches for patch correctness assessment.}
In this study, we adopt two patch datasets based on the recent studies \cite{tian2020evaluating,wang2020automated,lin2021context}, a small one containing 1,183 Defects4J labeled patches and a large one containing 50,794 real-world labeled patches.

\begin{table}[htbp]
  \centering
  \caption{\newrevise{Statistics of patches in the small dataset}}
    \begin{tabular}{l|rr|r}
    \toprule
    \newrevise{\#Tools} & \newrevise{\#Overfitting} & \newrevise{\#Correct} & \newrevise{\#Total} \\
    \midrule
    \newrevise{Nopol} & \newrevise{29} & \newrevise{1} & \newrevise{30} \\
    \newrevise{jGenProg} & \newrevise{16} & \newrevise{2} & \newrevise{18} \\
    \newrevise{GenProg} & \newrevise{24} & \newrevise{1} & \newrevise{25} \\
    \newrevise{SketchFix} & \newrevise{7} & \newrevise{5} & \newrevise{12} \\
    \newrevise{defects4j-dev} & \newrevise{0} & \newrevise{354} & \newrevise{354} \\
    \newrevise{DynaMoth} & \newrevise{21} & \newrevise{1} & \newrevise{22} \\
    \newrevise{CapGen} & \newrevise{41} & \newrevise{9} & \newrevise{50} \\
    \newrevise{TBar} & \newrevise{33} & \newrevise{7} & \newrevise{40} \\
    \newrevise{SOFix} & \newrevise{1} & \newrevise{10} & \newrevise{11} \\
    \newrevise{FixMiner} & \newrevise{19} & \newrevise{6} & \newrevise{25} \\
    \newrevise{ACS} & \newrevise{5} & \newrevise{16} & \newrevise{21} \\
    \newrevise{Kali} & \newrevise{36} & \newrevise{2} & \newrevise{38} \\
    \newrevise{kPAR} & \newrevise{32} & \newrevise{2} & \newrevise{34} \\
    \newrevise{Jaid} & \newrevise{40} & \newrevise{32} & \newrevise{72} \\
    \newrevise{AVATAR} & \newrevise{37} & \newrevise{17} & \newrevise{54} \\
    \newrevise{SequenceR} & \newrevise{45} & \newrevise{10} & \newrevise{55} \\
    \newrevise{Arja} & \newrevise{49} & \newrevise{8} & \newrevise{57} \\
    \newrevise{RSRepair} & \newrevise{31} & \newrevise{2} & \newrevise{33} \\
    \newrevise{ICSE18} & \newrevise{99} & \newrevise{28} & \newrevise{127} \\
    \newrevise{jKali} & \newrevise{18} & \newrevise{4} & \newrevise{22} \\
    \newrevise{SimFix} & \newrevise{42} & \newrevise{16} & \newrevise{58} \\
    \newrevise{jMutRepair} & \newrevise{14} & \newrevise{2} & \newrevise{16} \\
    \newrevise{Cardumen} & \newrevise{9} & \newrevise{0} & \newrevise{9} \\
    \midrule
    \newrevise{Total} & \newrevise{648} & \newrevise{535} & \newrevise{1183} \\
    \bottomrule
    \end{tabular}%
  \label{tab:small_dataset}%
\end{table}%

On the small dataset, we mainly focus on the released patches from Defects4J \cite{just2014defects4j}, which is the most widely-adopted benchmark in APR research \cite{liu2020efficiency}.
We select the \delete{benchmarks} \revise{datasets} released by two recent large-scale studies, i.e., Wang et al. \cite{wang2020automated} and Tian et al. \cite{tian2020evaluating}. 
Specifically, the first benchmark \cite{wang2020automated} includes the labeled patches provided by Liu et al. \cite{liu2020efficiency}, Xiong et al. \cite{xiong2018identifying} and Defects4J developers \cite{just2014defects4j}.
The second benchmark  \cite{tian2020evaluating} includes the labeled patches from Liu et al. \cite{liu2020efficiency} and also considers the patches generated by some well-known APR tools that are not included in Liu et al. \cite{liu2020efficiency} to better explore the overfitting problem, i.e., JAID \cite{2017Chen}, SketchFix \cite{hua2018towards}, CapGen \cite{wen2018context}, SOFix \cite{liu2018mining} and SequenceR \cite{chen2019sequencer}.
To avoid the data leakage issue in the two benchmarks, a filtering process is also conducted to discard duplicate patches.
In particular, given a patch whose all the blank spaces are removed, the left text information is compared with that from the other patches.
If two patches are identical concerning their text information, they are considered duplicates, resulting in 1,183 patches in our small dataset.
The patches are generated by 22 distinct APR tools, which can be divided into four categories, i.e., heuristic-based, constraint-based, template-based, and learning-based techniques.
The detailed information on these covered APR tools is presented in Table \ref{tab:small_dataset}, where the first column lists the four repair technique categories and the second column list the corresponding repair techniques.

\begin{table}[t]
\caption{Datasets used in our experiment}
\label{tab:dataset}
\begin{tabular}{l|l|rrr}
\toprule
Datasets & Subjects & \# Correct & \# Overfitting & Total \\
\midrule
\multirow{3}{*}{Small} & Tian et al. \cite{tian2020evaluating} & \newrevise{468} & \newrevise{532} & \newrevise{1,000} \\
 & Wang et al. \cite{wang2020automated} & \newrevise{248} & \newrevise{654} & \newrevise{902} \\
\cmidrule{2-5}
 & Our Study & \newrevise{535} & \newrevise{648} & \newrevise{1,183} \\
\midrule
\multirow{3}{*}{Large} & ManySStuBs4J \cite{karampatsis2020often} & \newrevise{51,433} & \newrevise{0} & \newrevise{51,433} \\
 & RepairThemAll \cite{durieux2019empirical} & \newrevise{900} & \newrevise{63,393} & \newrevise{64,293} \\
 \cmidrule{2-5}
 & Our Study & \newrevise{25,589} & \newrevise{24,105} & \newrevise{49,694} \\
\bottomrule
\end{tabular}
\end{table}

On the large dataset, we further consider a variety of patches generated from other benchmarks, to evaluate the generality of {\toolname}.
Recently, existing studies demonstrate that APR techniques may overfit Defects4J in terms of repairability \cite{durieux2019empirical,zhang2022program}.
Thus, some other benchmarks have been \delete{conducted} \revise{applied} to evaluate the performance of APR techniques, such as Bugs.jar \cite{2018Saha}, IntroclassJava \cite{durieux2016introclassjava}, BEARS \cite{2019Madeiral} and QuixBugs \cite{lin2017quixbugs}, providing substantial patches on the large dataset.
In this work, we consider a large patch dataset released by a recent study \cite{lin2021context} to investigate the generality of {\toolname}.
The large patch dataset includes the labeled patches provided from RepairThemAll framework \cite{durieux2019empirical} and ManySStuBs4J \cite{karampatsis2020often}.
In particular, RepairThemAll framework \cite{durieux2019empirical} contains 64,293 patches using 11 Java test-suite-based repair tools and 2,141 bugs from five diverse benchmarks.
\delete{However, there exists an imbalanced dataset issue as over 98.6\%\footnote{The RepairThemAll Framework. \url{https://github.com/program-repair/RepairThemAll}, accessed March 2023} (63,393/64,293) of generated patches are actually labeled as incorrect.}
\revise{However, Tian et al. \cite{tian2020evaluating} manually inspect the correctness of the patches from  RepairThemAll\footnote{The RepairThemAll Framework. \url{https://github.com/program-repair/RepairThemAll}, accessed March 2023} and find 98.6\% of patches (63,393/64,293) are actually overfitting ones, resulting in an imbalanced dataset.}
Recent studies have revealed that a well-balanced dataset is essential when investigating deep learning-based prediction techniques \cite{tian2020evaluating, ye2021automated}.
To compensate the lack of correct patches, the large patch dataset then includes ManySStuBs4J  \cite{karampatsis2020often}, which provides simple bug-fix changes mined from 1,000 popular open-source Java projects.
The bug-fix changes are correct fix attempts of real-world bugs and thus are considered correct patches in our experiment.
Finally, a large balanced patch dataset is built from the RepairThemAll framework and ManySStuBs4J by discarding duplicate patches and filtering the ones from small student-written programming assignments (e.g., IntroClassJava).
The dataset involves all available patches generated on RepairThemAll framework and ManySStuBs4J, resulting in 49,694 patches after deduplication.

Statistics on the two datasets are shown in Table \ref{tab:dataset}.
Table \ref{tab:dataset} has two main rows representing the two datasets, each of which has three sub-rows.
The first and second sub-rows list the two sources in the corresponding dataset.
The third column lists the filtered patches used in our experiment from the two sources.
We also present the number of correct, overfitting and total patches in the last three columns.

\subsection{Baselines}
\label{sec:baselines}

\begin{table*}[htbp]
  \centering
  \caption{\newrevise{Compared APCA techniques in our experiment}}
    \begin{tabular}{lll}
    \toprule
          & with Oracle Required & without Oracle Required \\
    \midrule
    Dynamic-based & Evosuite \cite{fraser2011evosuite}, Randoop \cite{pacheco2007randoop}, DiffTGen \cite{xin2017difftgen}, Daikon \cite{yang2020daikon} & PATCH-SIM \cite{xiong2018identifying}, E-PATCH-SIM \cite{xiong2018identifying},
R-Opad \cite{yang2020daikon}, E-Opad \cite{yang2020daikon} \\
    \midrule
    Static-based & $\bigotimes$  & ssFix \cite{xin2017ssfix}, CapGen \cite{wen2018context},
Anti-patterns \cite{tan2016anti}, S3 \cite{le2017s3} \\
    \midrule
    \multirow{2}{*}{Learning-based} & \multirow{2}{*}{$\bigotimes$}  & ODS \cite{ye2021automated}, Random Forest \cite{tian2020evaluating}, \\
     &   & \colorbox{lightgray}{Embedding learning \cite{tian2020evaluating}, CACHE \cite{lin2021context}, Our proposed {\toolname}} \\
    \bottomrule
    \multicolumn{3}{l}{\small \colorbox{lightgray}{~~} denotes the representation learning techniques.}\\
    \label{tab:baselines}%
    \end{tabular}%
\end{table*}%

Various APCA techniques have been proposed in the literature to validate patch correctness.
Following existing studies \cite{lin2021context,xiong2018identifying}, we attempt to select state-of-the-art techniques designed for Java language as Java is the most targeted language in APR community \cite{liu2020efficiency} and the existing patches of real-world bugs are usually available in Java language \cite{tian2020evaluating}.
We first consider the recent empirical study by Wang et al. \cite{wang2020automated} to identify existing APCA techniques.
We then select recent advanced studies \cite{lin2021context,tian2020evaluating} that are not included in Wang et al. \cite{wang2020automated}.

In general, following existing work \cite{lin2021context,wang2020automated,zhou2023patchzero,le2023invalidator}, the existing APCA techniques can be categorized into static, dynamic and learning-based APCA techniques according to whether test execution is needed or deep learning techniques are adopted (mentioned in Section \ref{sec:bg&mv}). 
Meanwhile, according to whether the ground-truth patch is required, they can be further categorized into two categories (i.e., with or without oracle).
Particularly, similar to our proposed method {\toolname}, \delete{CHCHE} \revise{CACHE} and embedding learning techniques adopt representation models to embed changed code and a deep learning classier to predict patch correctness.
Such techniques can be further considered as representation learning APCA techniques.

The details of the selected APCA techniques are illustrated in Table \ref{tab:baselines}.
The first column lists three APCA categories.
The second and third columns list whether the oracle information is equipped.
We also list the representation learning techniques (e.g., {\toolname}) in the light gray box.
We summarize the selected techniques as follows.

\subsubsection{Dynamic-based APCA Techniques}
Dynamic-based techniques are designed to distinguish correct patches from overfitting patches based on the outcome or the execution traces of the original or generated test cases.

\textbf{\textit{Simple Test Generation}}.
The overfitting issue is prevalent in the repair process due to the weak adequacy of existing test cases.
Thus, researchers use test case generation tools to generate extra test cases based on the fixed program and check whether or not the generated patches that pass the original test cases can pass the extra test cases \cite{ye2021scale, smith2015cure}.
In this work, we adopt Evosuite \cite{fraser2011evosuite} and Randoop \cite{pacheco2007randoop} as the test case generation tools, as they have been widely investigated in previous studies.

\textbf{\textit{DiffTGen}}.
Xin et al. \cite{xin2017difftgen} identify overfitting patches by executing test cases generated by an external test generator (i.e., Evosuite).
Different from \textit{simple test generation} generating test cases randomly, DiffTGen generates test cases to uncover the syntactic differences between the patched and buggy program.
A plausible patch is regarded as overfitting if the output of the patched program is not the same as that of the correct program.
DiffTGen needs a human-written patch as a reference and requires providing human-amenable testing information for the developers to provide oracles the generated test cases.

\textbf{\textit{Daikon}}.
Daikon is a dynamic-based technique based on the program invariant with oracle information.
Yang et al. \cite{yang2020daikon} adopt the program invariant to explore the differences between an overfitting and a correct patch.
A patch is considered correct if its inferred invariant is identical to that of the ground-truth.
If there exists a different comparison, the patch is considered overfitting.

\textbf{\textit{PATCH-SIM}}.
Xiong et al. \cite{xiong2018identifying} consider the execution traces of the passing tests on the buggy and patched programs are likely to be similar, while the execution traces of failing tests on the buggy and patched programs are likely different.
Based on the concept, they approximate the correctness of a patch based on the execution trace without the oracle information.
PATCH-SIM adopts Randoop to generate additional test cases to collect dynamic execution information.
In this work, we also replace Randoop with Evosuite to comprehensively explore the impact of test generation techniques (denoted as E-PATCH-SIM).

\textbf{\textit{Opad}}.
Yang et al. \cite{yang2017better} adopt fuzzing testing to generate new test cases and employ two test oracles (crash and memory-safety) to enhance the validity checking of patches.
The original implementation of Opad is not designed for Java language and uses American Fuzz Lop (AFL) as the fuzzing technique.
In this work, following recent studies \cite{wang2020automated, lin2021context}, we replace AFL with Randoop and Evosuite to generate new test cases on the Java programs and denote them as R-Opad and E-Opad, respectively.

 \subsubsection{Static-based APCA Techniques}
 Static-based techniques usually adopt static analysis tools to extract some designed static features and then check patch correctness based on such features.

\textbf{\textit{ssFix}}.
ssFix \cite{xin2017ssfix} is a static-based technique that utilizes token-based syntax representation to generate patches with a higher probability of correctness.
ssFix first performs a syntactic code search to find code snippets from a codebase that is syntax-related to the context of a bug to generate correct patches, and then prioritizes the patches based on the modification types and the modification sizes.

\textbf{\textit{CapGen}}.
Wen et al. \cite{wen2018context} propose three aspects of context information (i.e., genealogy contexts, variable contexts and dependency contexts) embedded in an AST node and its surrounding codes to prioritize correct patches over overfitting ones.
In this work, following recent studies \cite{wang2020automated, lin2021context}, we extract the three context information as static features to investigate patch correctness assessment.

\textbf{\textit{Anti-patterns}}.
Tan et al. \cite{tan2016anti} define a set of rules that essentially capture disallowed modifications to the buggy program, and a patch is overfitting if it falls into the rules.
A recent study \cite{wang2020automated} has shown that the manually-defined anti-patterns may have false positives for correct patches, resulting in destructive effects in patch correctness prediction.

\textbf{\textit{S3}}.
Le et al. \cite{le2017s3} assume that a correct patch is often syntactically and semantically close to a buggy code snippet.
Thus, they adopt six syntactic features (i.e., AST differencing, cosine similarity and locality of variables and constants) and semantic features (i.e., model counting, output coverage and anti-patterns) to measure the distance between a candidate patch and the buggy code snippet.

 \subsubsection{Learning-based APCA Techniques}
\delete{
Learning-based techniques can predict whether a plausible patch is correct or not based on machine learning techniques.}
\revise{Learning-based techniques mainly investigate static features and machine learning techniques to build predictive models for patch correctness prediction.}

\textbf{\textit{ODS}}.
Ye et al. \cite{ye2021automated} first extract 202 code features at the abstract syntax tree level and then use supervised learning to learn a probabilistic model automatically.
The results show that ODS can achieve better prediction performance than the dynamic-based technique PATCH-SIM with a faster speed.

\textbf{\textit{CACHE}}.
Lin et al. \cite{lin2021context} propose a context-aware APCA technique CACHE by taking both the changed code snippet and the correlated unchanged code snippet into consideration.
CACHE first parses the patched code snippet into AST representation and then utilizes the AST path technique to capture the structure information.

\textbf{\textit{Random Forest}}.
Wang et al. \cite{wang2020automated} investigate the effectiveness of adopting deep learning models to predict patch correctness based on eight static features (two from ssFix, three from S3, and three from CapGen).
To integrate the static features, six widely-used classification models (including Random Forest, Decision Table, J48, Naive Bayes, Logistic Regression, and SMO) are adopted.
The results demonstrate that Random Forest can achieve both superior precision and recall performance.
In this work, following existing work \cite{lin2021context}, we also adopt Random Forest to predict the patch correctness based on the integrated static features.

\textbf{Embedding Learning}.
Tian et al. \cite{tian2020evaluating} propose to leverage representation learning techniques to produce embedding for buggy and patched code snippets and then adopt supervised learning classifies to predict patch correctness.
In particular, nine representation learning APCA techniques are evaluated, involving three embedding techniques (i.e., CC2vec, BERT and Doc2Vec) and three classifiers (logistic regression, decision tree and naive bayes).
\revise{
{\toolname} differs from Tian et al. \cite{tian2020evaluating} in that BERT is adopted as one of the embedding techniques to embedding code without any training, while we attempt to take advantage of the generic knowledge of the pre-train model and the classifier by further fine-tuning {\toolname} to support patch correctness assessment.
Considering the fact that off-the-shelf pre-trained models (e.g., BERT) are pre-trained with data corpora in other fields (e.g., NLP) and thus may not be suitable to embed patched code snippets, fine-tuning pre-trained model in our APPT architecture can obtain optimal embedding vectors for reasoning \newrevise{about} patch correctness.
}

\subsection{Model Selection}
To the best of our knowledge, {\toolname} is the first automated patch correctness prediction technique by fine-tuning the existing pre-trained model \revise{in the APCA domain}.
In this paper, we adopt BERT as the encoder stack due to its powerful performance in previous work \cite{devlin2018bert}.

Specifically, BERT is pre-trained on large amounts of text data with two self-supervised goals, i.e., masked language modeling (MLM) and next sentence prediction (NSP).
MLM aims to let the model predict the masked words by masking 15\% of words in each sentence randomly.
NSP aims to further improve the model’s ability to understand the relationship between two sentences by letting the model predict whether the given sentence pair is continuous.
The model then can be fine-tuned to adapt to some specific downstream tasks and has achieved remarkable state-of-the-art results on a variety of natural language processing tasks, such as question answering and language inference.

There exist two model architectures at different sizes, i.e., BERT$_{base}$ and BERT$_{large}$ \cite{devlin2018bert}.
The former has 12 layers and 12 attention heads, and the embedding size is 768, while the latter has a double layer number and 16 attention heads, and the embedding size is changed to 1024.
In this paper, we do not modify the vocabulary size and use the pre-trained BERT$_{base}$ as the fine-tuning starting point instead of starting from scratch.

In this paper, {\toolname} is conceptually and practically generalizable to various pre-trained models.
We also select CodeBERT and GraphCodeBERT as the encoder stack to evaluate the scalability of {\toolname}.
CodeBERT and GraphCodeBERT share the same model architecture as BERT, while utilizing paired natural language and programming language to pre-train the model to support code-related tasks (mentioned in Section \ref{sec:se}).

\subsection{Evaluation Metrics}
\label{sec:metric}
We evaluate the prediction performance of various APCA approaches by accuracy, precision, recall, F1-score and AUC metrics, which have been widely adopted in patch correctness assessment research and other classification tasks \cite{lin2021context,tian2020evaluating}.
Given the number of 
true positives (TPs, a TP refers to an overfitting patch that is identified as overfitting), 
false positives (FPs, a FP refers to a correct patch that is identified as overfitting), 
false negatives (FNs, a FN refers to an overfitting patch is identified as correct) 
and true negatives (TNs, a TN refers to a correct patch that is identified as correct), 
the metrics are defined as follows:

\emph{$\bullet$ Accuracy:} the proportion of correctly reported (whether the patch is correct or not) patches.
Accuracy measures the probability that the prediction of APCA techniques is correct.

\begin{equation}
    Accuracy = \frac{TP+TN} {TP+FP+FN+TN} \\
\end{equation}

\emph{$\bullet$ Precision:} the proportion of real overfitting patches over the reported overfitting patches.
Precision measures how much we can trust the APCA techniques when it predicts a patch as overfitting.

\begin{equation}
    Precision =  \frac{TP} {TP+FP} \\
\end{equation}

\emph{$\bullet$ Recall:} the proportion of reported overfitting patches over all the real overfitting patches.
Recall measures the ability of the APCA techniques to find all the overfitting patches in the dataset.

\begin{equation}
    Recall = \frac{TP} {TP+FN}
\end{equation}

\emph{$\bullet$ F1-score:} twice the multiplication of precision and recall divided by the sum of them.
F1-score measures the trade-off between precision and recall by taking their harmonic mean.

\begin{equation}
    F1\text{-}score = 2 * \frac{Precision * Recall }{ Precision + Recall}
\end{equation}

\emph{$\bullet$ AUC:} the entire two-dimensional area underneath the entire receiver operating characteristic curve.
AUC measures the probability that the classifier will rank a randomly chosen overfitting patch higher than that of a randomly chosen correct patch. 
The higher the AUC, the better the APCA techniques is at predicting real overfitting patches as overfitting and real correct patches as correct.

\delete{
\begin{equation}
\begin{gathered}
AUC = \frac{\sum I\left(P_{\text{overfitting }}, P_{\text{correct }}\right)}{M \times N} \\
I\left(P_{\text{overfitting}}, P_{\text{correct} }\right)=\left\{\begin{array}{l}
1, P_{\text{overfitting}}>P_{\text{correct}} \\
0.5, P_{\text{overfitting}}=P_{\text{correct}} \\
0, P_{\text{overfitting}}<P_{\text{correct}}
\end{array}\right.
\end{gathered}
\end{equation}}
\revise{
\begin{equation}
AUC=\frac{\sum_{\mathrm{i} \in \text{O}} \operatorname{rank}_i - M\left(M+1\right) / 2}{M \times N}
\end{equation}}
\delete{where $M$ and $N$ denote the number of overfitting and correct patches, while $P_{\text{overfitting}}$ and $P_{\text{correct}}$ denote the prediction probability for the overfitting and correct patches.}
\revise{
where $M$ and $N$ denote the number of overfitting and correct patches, respectively.
$O$ denotes the overfitting patch set and ${rank}_i$ denotes the rank of the $i-$th overfitting patch in the descending list of output probability produced by each model.
}

\subsection{Implementation Details}
All of our approaches are built based on PyTorch framework\footnote{PyTorch. \url{https://pytorch.org/}, accessed March 2023}.
We use the Hugging Face\footnote{Hugging Face. \url{https://huggingface.co/}, accessed March 2023}
implementation version of BERT in our work.
Considering previous work recommendation~\cite{raffel2019t5,yuan2022circle},  we utilize ``bert-base-uncased'' (refer to BERT$_{base}$) as the initial point,  as the base version is quite lightweight to employ in practice with comparable effectiveness compared against the large version.
There exist $12$ layers of transformer blocks and $12$ self-attention heads in the ``bert-base-uncased'' model.
The optimizer is Adam~\cite{kingma2014adam} with $5e-5$ learning rate.
The batch size is $16$ and dropout rate is $0.5$.
We train for most $50$ epochs and the max length of the input is set to $512$ due to model limitation.

All the training and evaluation of our methods are conducted on one Ubuntu 18.04.3 server with two Tesla V100-SXM2 GPUs.

\section{Results and Analysis}
\label{sec:re&an}

\subsection{RQ1: \delete{Comparing with Representation Learning-based Techniques} \revise{Effectiveness of {\toolname}}}
\label{sec:rq1}

\begin{table*}[htbp]
\centering
\caption{Effectiveness of {\toolname} compared with representation learning-based APCA techniques on the small dataset}
\label{tab:rq1_small_dataset}

\begin{tabular}{c|c|ccccc}
\toprule
Classifier & Embedding & Accuracy & Precision & Recall & F1-score & AUC \\
\midrule
\multirow{5}{*}{Decision Tree} & BERT & 63.5\% & 65.3\% & 70.9\% & 67.9\% & 63.7\% \\
 & CC2vec & 66.1\% & 69.4\% & 68.0\% & 68.7\% & 66.5\% \\
 & code2vec & 65.1\% & 68.1\% & 68.3\% & 68.1\% & 64.4\% \\
 & code2seq & 60.1\% & 63.5\% & 64.0\% & 63.7\% & 60.0\% \\
 & Doc2Vec & 61.2\% & 64.5\% & 65.3\% & 64.8\% & 60.8\% \\
 \midrule
\multirow{5}{*}{Logistic Regression} & BERT & 64.8\% & 66.5\% & 72.4\% & 69.2\% & 68.7\% \\
 & CC2vec & 64.9\% & 62.4\% & 90.1\% & 73.7\% & 68.6\% \\
 & code2vec & 66.8\% & 68.6\% & 72.9\% & 70.6\% & 70.2\% \\
 & code2seq & 60.7\% & 63.3\% & 67.6\% & 65.3\% & 63.1\% \\
 & Doc2Vec & 63.7\% & 65.7\% & 70.8\% & 68.0\% & 68.9\% \\
 \midrule
\multirow{5}{*}{Naïve Bayes} & BERT & 61.6\% & 64.8\% & 65.7\% & 65.0\% & 64.7\% \\
 & CC2vec & 60.0\% & 58.3\% & \textbf{94.6\%} & 72.2\% & 58.1\% \\
 & code2vec & 57.7\% & 58.1\% & 81.5\% & 67.8\% & 55.6\% \\
 & code2seq & 57.0\% & 59.0\% & 70.5\% & 64.2\% & 60.6\% \\
 & Doc2Vec & 64.1\% & 65.8\% & 72.4\% & 68.7\% & 67.0\% \\
 \midrule
\multicolumn{2}{c|}{CACHE} & 75.4\% & 79.5\% & 76.5\% & 78.0\% & 80.3\% \\
\midrule
\multicolumn{2}{c|}{\toolname} & \textbf{79.7\%} & \textbf{80.8\%} & 83.2\% & \textbf{81.8\%} & \textbf{82.5\%} \\

\bottomrule
\end{tabular}
\end{table*}

\subsubsection{\revise{Comparing with Representation Learning-based Techniques}}

\emph{\textbf{Experimental Design.}}
As discussed in Section \ref{sec:baselines}, {\toolname}, CACHE and embedding learning techniques (i.e., techniques within the light gray box in Table \ref{tab:baselines}) can be categorized as representation learning APCA techniques.
In this section, we aim to explore the performance of {\toolname} when compared with these representation learning techniques.
In particular, embedding learning techniques \cite{tian2020evaluating} mainly adopt embedding models (i.e., BERT, Doc2Vec, and CC2Vec) to embed buggy and patched code fragments, and then train classification models (i.e., Decision Tree, Logistic Regression, and Naive Bayes) to predict patch correctness.
Following previous study \cite{lin2021context}, we also consider two additional embedding models (i.e., code2vec and code2seq) in the experiment.
Meanwhile, CACHE can also be considered as a representation learning technique, which incorporates the context information in embedding code changes, and trains a deep learning classifier to predict the patch correctness.

In total, \delete{16 representation learning techniques are considered in our experiment, involving five embedding techniques multiplied by three classification models, and one context-aware representation learning technique CACHE.}
\revise{one representation learning technique with 15 settings from Tian et al. \cite{tian2020evaluating} (involving five embedding techniques multiplied by three classification models), and one context-aware representation learning technique CACHE are considered in our experiment.}
Following the previous study \cite{tian2020evaluating}, we perform \revise{standard practice} 5-fold cross-validation on both the small and large datasets for comparison.

\emph{\textbf{Results.}}
Comparison results against the existing representation learning techniques are presented in Table \ref{tab:rq1_small_dataset} to Table \ref{tab:rq1_large_dataset} for both the small and large datasets.
The first column lists the three classifiers and the second column lists \delete{the five embedding approaches }\revise{the five off-the-shelf embedding models}.
The remaining columns list the detailed values of accuracy, precision, recall, F1-score and AUC metrics, respectively.
We present the most recent representation learning work CACHE and our {\toolname} in the bottom part of Table \ref{tab:rq1_small_dataset} and Table \ref{tab:rq1_large_dataset}.
It can be observed that {\toolname} achieves the best performance under each experimental setting.

On the small dataset, {\toolname} is around 
4.3\%, 1.3\%, 6.7\%, 3.8\% and 2.2\%
higher than the state-of-the-art technique CACHE in terms of all metrics (i.e., 79.7\% vs. 75.4\% for accuracy, 80.8\% vs. 79.5\% for precision, 83.2\% vs. 76.5\% for recall, 81.8\% vs. 78.0\% for F1-score, and 82.5\% vs. 80.3\% for AUC).
Compared with all representation learning techniques, {\toolname} achieves the best performance in terms of accuracy, precision, F1-score and AUC metrics.
In particular, the values of {\toolname} on the accuracy and precision metrics are 79.7\% and 80.8\%, respectively, while the optimal values of all other techniques are 75.4\% and 79.5\%. 
This suggests that {\toolname} can generally achieve the most accurate predictions, and the patches identified as overfitting by {\toolname} are of high confidence to be overfitting. 
Regarding recall, the values of CC2vec and code2vec can sometimes exceed those of {\toolname} since they tend to classify most patches as overfitting (e.g., CC2vec with Naive Bayes classifies 1,051 out of 1,183 patches as overfitting and thus achieves a high recall of 94.6\%).
However, these techniques achieve relatively low precision (e.g., CC2vec with Naive Bayes classifier has only 72.2\% for recall).
In contrast, {\toolname} can achieve a high recall exceeding 83\% while maintaining a high precision of 80.8\%.

\begin{table*}[htbp]
\centering
\caption{Effectiveness of {\toolname} compared with representation learning techniques on the large dataset}
\label{tab:rq1_large_dataset}
\begin{tabular}{c|c|ccccc}
\toprule
Classifier & Embedding & Accuracy & Precision & Recall & F1-score & AUC \\
\midrule
\multirow{5}{*}{Decision Tree} & BERT & 95.7\% & 93.9\% & 97.4\% & 95.6\% & 95.9\% \\
 & CC2vec & 95.6\% & 95.4\% & 95.7\% & 95.5\% & 95.7\% \\
 & code2vec & 95.0\% & 93.2\% & 96.6\% & 94.9\% & 95.4\% \\
 & code2seq & 92.2\% & 91.0\% & 93.2\% & 92.3\% & 92.4\% \\
 & Doc2Vec & 85.1\% & 84.2\% & 85.3\% & 84.7\% & 85.3\% \\
 \midrule
\multirow{5}{*}{Logistic Regression} & BERT & 82.4\% & 83.6\% & 79.4\% & 81.4\% & 91.0\% \\
 & CC2vec & 91.2\% & 96.1\% & 85.4\% & 90.4\% & 95.0\% \\
 & code2vec & 89.6\% & 88.6\% & 90.2\% & 89.4\% & 95.0\% \\
 & code2seq & 91.5\% & 90.5\% & 92.2\% & 91.4\% & 96.0\% \\
 & Doc2Vec & 90.4\% & 91.9\% & 88.0\% & 89.9\% & 96.1\% \\
 \midrule
\multirow{5}{*}{Naïve Bayes} & BERT & 68.2\% & 80.3\% & 45.7\% & 58.2\% & 74.6\% \\
 & CC2vec & 78.4\% & 94.8\% & 58.6\% & 72.5\% & 92.4\% \\
 & code2vec & 61.4\% & 68.7\% & 37.4\% & 48.4\% & 69.3\% \\
 & code2seq & 70.3\% & 76.8\% & 55.5\% & 64.5\% & 78.9\% \\
 & Doc2Vec & 81.2\% & 86.4\% & 75.5\% & 78.9\% & 88.9\% \\
 \midrule
\multicolumn{2}{c|}{CACHE} & 98.6\% & 98.9\% & 98.2\% & 98.6\% & 98.9\% \\
\midrule
\multicolumn{2}{c|}{\toolname} & \textbf{99.1\%} & \textbf{99.1\%} & \textbf{99.1\%} & \textbf{99.1\%} & \textbf{99.9\%} \\
\bottomrule
\end{tabular}
\end{table*}

On the large dataset, we can find {\toolname} achieves over 99\% for the five metrics, outperforming all existing approaches.
For example, {\toolname} reaches 99.9\% in terms of AUC, which is 1.0\% higher than the second highest value obtained from the most recent technique CACHE (i.e., 98.9\%).
This suggests that {\toolname} is more capable of distinguishing correct and overfitting patches than CACHE.
Besides, the improvement against CACHE for accuracy, precision, recall and F1-score metrics achieves 0.5\%, 0.3\%, 0.9\% and 0.5\%, respectively. 
We also find that the performance achieved on the large dataset is commonly higher than that achieved on the small dataset.
For example, the average value among the five metrics increases from 81.06\% to 99.26\%, resulting in a 22.5\% improvement rate.
Based on our analysis on the two datasets, the possible reason for this improvement is that bugs on the large dataset are usually simple.
We observe that all ManySStuBs4J patches on the large dataset are single-line operations, while patches on the small dataset usually cross multiple lines (e.g., more than 40\% of Defects4J developer patches are multiple line patches \cite{lin2021context}).
It is easy for neural networks to learn the correctness distribution of such simple code changes.
Meanwhile, the difference in patch scale between the two datasets may be the second reason.
We find there exist 49,694 patches on the large dataset, which is 42 times larger than that of the small dataset.
The amount of training data is often the single most dominant factor that determines the performance of the neural networks \cite{ye2022neural}.
More available patches benefit the neural networks to learn diverse code changes better.

\revise{
\textbf{Case study of unique identified overfitting and correct patch.}
Fig. \ref{fig:case} presents an example of an overfitting patch generated for bug Math-53, which is detected as overfitting by {\toolname} but not CACHE and Tian et al. \cite{tian2020evaluating}.
In this bug, the method \textit{add()} (line 1 in $P_3$) is used to return a Complex whose value is \textit{this} + \textit{rhs} (\textit{rhs} is the parameter value to be added to this Complex).
Ideally, if either \textit{this} or \textit{rhs} has a NaN value in either part, NaN is returned; otherwise Infinite and NaN values are returned in the parts of the result according to the rules for Double arithmetic.
In the buggy version, \textit{add()} directly creates a complex number given the real and imaginary parts.
Consequently, the buggy version fails to check whether or not the parameter has a NaN value and throws AssertionFailedError.
As we find in $P_1$, Jaid fixes the bug by adding the condition \textit{if ((isNaN() \(\vert \vert\) rhs.isNaN()) == true)} (line 4 in $P_1$), which is equivalent the condition written by developers \textit{if (isNaN \(\vert \vert\) rhs.isNaN)} (line 4 in $P_3$).
However, CACHE fails to detect the semantic equivalence between the developer patch and the correct patch.
In contrast, {\toolname}, which relies on a pre-trained model and fine-tuning process, still can correctly identify the correct patch.
Similarly, another example can be seen in $P_2$, in which an overfitting patch generated by HDRepair cannot be detected by CACHE but by {\toolname} as {\toolname} successfully captures the different behaviors between the expression \textit{rhs.getArgument()} (line 5 in $P_2$) and \textit{rhs.getArgument()} (line 5 in $P_3$).
}

\begin{figure}[htbp]
\centering
\ttfamily
\footnotesize
    \begin{tabular}{p{0.1cm}p{0.1cm}p{6.8cm}}

    \hline
    \multicolumn{3}{l}{\revise{\textbf{A Correct Patch $P_1$ Generated by Jaid}}} \\
        1 & & \revise{public Complex add(Complex rhs)} \\
        2 & & \qquad \revise{throws NullArgumentException \{} \\
        3 & & \qquad \revise{MathUtils.checkNotNull(rhs);} \\
        4 & \textcolor{blue}{+} & \qquad \revise{if((isNaN() || rhs.isNaN()) == true) \{} \\
        5 & \textcolor{blue}{+} & \qquad \qquad \revise{return NaN;} \\
        6 & \textcolor{blue}{+} & \qquad \revise{\}} \\
        7 & & \qquad \revise{return createComplex(real + rhs.getReal(),} \\
        8 & & \qquad \qquad \revise{imaginary + rhs.getImaginary());} \\
        9 & & \} \\
    \hline
    \multicolumn{3}{l}{\revise{\textbf{A Plausible Patch $P_2$ Generated by HDRepair}}} \\
        1 & & \revise{public Complex add(Complex rhs)} \\
        2 & & \qquad \revise{throws NullArgumentException \{} \\
        3 & & \qquad \revise{MathUtils.checkNotNull(rhs);} \\
        4 & \textcolor{red}{-} & \qquad \revise{return createComplex(real + rhs.getReal(),} \\
        5 & \textcolor{blue}{+} & \qquad \revise{return createComplex(real + rhs.getArgument(),} \\
        6 & & \qquad \qquad \revise{imaginary + rhs.getImaginary());} \\
        7 & & \} \\
    \hline
    \multicolumn{3}{l}{\revise{\textbf{A Correct Patch $P_3$ Generated by Developers}}} \\
        1 & & \revise{public Complex add(Complex rhs)} \\
        2 & & \qquad \revise{throws NullArgumentException \{} \\
        3 & & \qquad \revise{MathUtils.checkNotNull(rhs);} \\
        4 & \textcolor{blue}{+} & \qquad \revise{if (isNaN || rhs.isNaN) \{} \\
        5 & \textcolor{blue}{+} & \qquad \qquad \revise{return NaN;} \\
        6 & \textcolor{blue}{+} & \qquad \revise{\}} \\
        7 & & \qquad \revise{return createComplex(real + rhs.getReal(),} \\
        8 & & \qquad \qquad \revise{imaginary + rhs.getImaginary());} \\
        9 & & \revise{\}} \\

    \hline
    \end{tabular}
\caption{\revise{APR-generated and developer patches for Math-53}}
\label{fig:case}
\end{figure}

\finding{1.1}{Overall, our analysis on representation learning techniques reveals that
(1) {\toolname} can outperform a state-of-the-art representation learning technique CACHE under all metrics and datasets.
(2) on the small dataset, {\toolname} achieves 79.7\% for accuracy and 82.5\% for AUC, which surpass CACHE by 4.3\% and 2.2\%.
(3) on the large dataset, {\toolname} exceeds 99\% on all metrics, yet none of existing representation learning techniques achieves that.}

\subsubsection{Comparing with APCA Techniques}

 \emph{\textbf{Experimental Design.}}
In this section, we aim to further compare the proposed method {\toolname} with the existing APCA techniques.
We select the remaining techniques mentioned in Section \ref{sec:baselines} (except representation learning techniques discussed in RQ1).
In total, 14 APCA techniques are considered in the experiment, involving four static techniques (Anti-patterns, ssFix, CapGen and S3), eight dynamic techniques (Evosuite, Randoop, DiffTGen, Daikon, R-Opad, E-Opad, PATCH-SIM and E-PATCH-SIM) and two learning techniques (Random Forest and ODS).

As it is time-consuming to run all the techniques (especially for dynamic and learning ones), following the existing work \cite{lin2021context}, we reuse the released results from the recent work \cite{wang2020automated, ye2021automated, lin2021context}.
We collect the detailed results of all selected APCA techniques from Lin et al. \cite{lin2021context}, which are concluded based on 902 patches (i.e., Wang et al. \cite{wang2020automated} in Table \ref{tab:dataset}) and a 10-fold cross-validation.
To fairly compare with all the state-of-the-art techniques, we perform our experiment in the same experimental setting.

\begin{table*}[htbp]
\centering
\caption{Effectiveness of {\toolname} compared with the traditional and learning-based APCA technique}
\label{tab:rq2_effectiveness}
\begin{tabular}{c|c|c|cccc}
\toprule
\multicolumn{2}{c|}{Category} & APCA & Accuracy & Precision & Recall. & F1-score \\
\midrule
\multirow{8}{*}{Dynamic-based} & \multirow{4}{*}{\rotatebox{90}{{w-oracle}}} & Evosuite & 65.9\% & 99.1\% & 53.5\% & 69.5\% \\
 & & Randoop & 51.3\% & 97.4\% & 33.8\% & 50.2\% \\
 & & DiffTGen & 49.6\% & 97.4\% & 30.6\% & 46.6\% \\
 & & Daikon & 76.1\% & 89.9\% & 73.7\% & 81.0\% \\
\cmidrule(lr){2-7}
 & \multirow{4}{*}{\rotatebox{90}{{wo-oracle}}} & R-Opad & 34.9\% & \textbf{100.0\%} & 10.2\% & 18.5\% \\
 & & E-Opad & 37.7\% & \textbf{100.0\%} & 14.7\% & 25.6\% \\
 & & PATCH-SIM & 49.5\% & 83.0\% & 38.9\% & 53.0\% \\
 & & E-PATCH-SIM & 41.7\% & 82.1\% & 25.8\% & 39.3\% \\
 \midrule
\multicolumn{2}{c|}{\multirow{4}{*}{Static-based}} & Anti-patterns & 47.6\% & 85.5\% & 33.5\% & 48.1\% \\
\multicolumn{2}{c|}{} & S3 & 69.7\% & 79.3\% & 78.9\% & 79.0\% \\
\multicolumn{2}{c|}{} & ssFix & 69.2\% & 78.9\% & 78.8\% & 78.8\% \\
\multicolumn{2}{c|}{} & CapGen & 68.0\% & 78.3\% & 77.4\% & 77.8\% \\
 \midrule
\multicolumn{2}{c|}{\multirow{2}{*}{Learning-based}} & Random Forest & 72.5\% & 87.0\% & 89.1\% & 88.0\% \\
 \multicolumn{2}{c|}{} & ODS & 88.9\% & 90.4\% & 94.8\% & 92.5\% \\
\midrule
\multicolumn{3}{c|}{\toolname} & \textbf{90.4\%} & 91.5\% & \textbf{96.0\%} & \textbf{93.6\%} \\
 
\bottomrule
\end{tabular}
\end{table*}

\emph{\textbf{Results.}}
The experiment results are listed in Table \ref{tab:rq2_effectiveness}.
The first two columns list the selected techniques and their corresponding categories.
The remaining columns list the detailed values of accuracy, precision, recall and F1-score metrics.

Compared with traditional dynamic-based and static-based APCA techniques, we can find that {\toolname} reaches 90.4\%, 96.0\% and 93.6\% in terms of accuracy, recall and F1-score, respectively.
Specifically, {\toolname} achieves the best overall performance with the three metrics, and none of the previous techniques exceeds 90\%.
As for precision, more than 91\% of patches reported by {\toolname} are indeed overfitting patches, which is better than all static-based techniques and three dynamic-based techniques (i.e., Daikon, PATCH-SIM, and E-PATCH-SIM).
Although some dynamic ones have higher precision values, it is time-consuming to generate additional test cases and collect run-time information.
More importantly, the recall of these techniques is usually low (e.g., 10.3\% for R-Opad), or the ground-truth oracle is needed (e.g., Evosuite and Randoop techniques), limiting the application of such techniques in practice.

Compared with learning-based techniques, we find that {\toolname} still performs better than a state-of-the-art technique ODS with respect to all four metrics (90.4\% vs. 88.9\% for accuracy, 91.5\% vs. 90.4\% for precision, 96.0\% vs. 94.8\% for recall, 93.6\% vs. 92.5\% for F1-score, respectively).
Overall, the improvement against Random Forest and ODS reaches 4.5\%$\sim$17.9\% and 1.1\%$\sim$1.5\%.
Considering that it is expensive for ODS to extract hundreds of manually-designed code features at AST level, our approach simply adopting the pre-trained model to encode a sequence of tokens is even more promising. 
We also highlight this direction of integrating code-aware features (e.g., code edits and AST representation) with pre-trained models for patch correctness assessment.

\finding{1.2}{Overall, our comparison results reveal that, 
(1) {\toolname} can achieve remarkable performance compared to exiting static-based techniques with a high recall reaching 96.0\%;
(2) {\toolname} can achieve higher precision than a state-of-the-art dynamic-based technique PATCH-SIM by 8.5\%;
(3) compared with existing learning-based techniques, {\toolname} can achieve the best performance among all metrics.
}
\subsection{RQ2: The Impact Analysis}
\revise{In this section, we further explore how different experiment choices affect the prediction performance of {\toolname}.}

\delete{
To further explore how different fine-tuning choices affect the prediction performance of pre-trained models, we first consider and replace the head-only token truncation with other truncation methods, such as  hybrid, mid-only, and tail-only token truncation.
We then adopt different methods to merge the buggy method vector and patched method vector, such as concatenate, additional, subtraction, and product operation.
We also mix the above-mentioned merged vectors as an additional concatenation method \revise{(detailed in Section \ref{sec:classification})}.
Recently, following the BERT model architecture, researchers use some code-related pre-trained tasks to capture the semantic connection between natural language and programming language, so as to further adapt these pre-training models for programming language.
Thus, we replace the BERT with two advanced models pre-trained with the programming language, i.e., CodeBERT \cite{feng2020codebert} and GraphCodeBERT \cite{guo2020graphcodebert}.
}

\subsubsection{The impact of Training Choice}

\revise{
\emph{\textbf{Experimental Design.}}
{\toolname} employs a pre-trained language model as the encoder stack, which is connected with an LSTM stack for classification training.
The quality of vector representation heavily relies on the language models of code being used.
In this process, the pre-trained model is further fine-tuned to obtain a suboptimal vector representation of code for patch correctness assessment.
Thus, we formulate this subRQ to investigate the impact of the pre-training, fine-tuning, and LSTM components.
}

\revise{
\emph{\textbf{Results.}}
Table \ref{tab:training} presents the ablation study results under different training choices.
The first column lists the two datasets.
The second column lists the three training choices, i.e., without pre-training, fine-tuning, and LSTM components.
The remaining columns list the detailed values of accuracy, precision, recall and F1-score and AUC metrics.}

\revise{
Generally speaking, all training components make contributions to the performance of {\toolname} in terms of these metrics.
For example, if the LSTM stack is not included on the small dataset, the accuracy and recall of {\toolname} will be decreased by 1.64\% and 7.02\%.
This finding demonstrates the rationale of our motivation that the LSTM stack is suitable to extract more hidden code change features for patch correctness assessment.}

We find that the fine-tuning component contributes the most to the overall performance of {\toolname} without which the Precision will degrade the most for both datasets.
For instance, if we do not fine-tune the pre-trained model, the precision of {\toolname} will be decreased by 13.16\% on the small dataset. 
This finding demonstrates the rationale of our motivation that fine-tuning the pre-trained model can help better convert the code snippets into the embedding, which is quite different from Tian et al. \cite{tian2020evaluating}.
We also note if we do not employ the pre-training knowledge (i.e., training the pre-trained model from scratch), the performances will drop at notable degrees (e.g., 7.26\% and 11.56\% for accuracy and precision on the small dataset).
This finding highlights the substantial benefits of the pre-training process on the larger corpus to assess the correctness of a patch.

\newrevise{
When comparing the improvements in the two datasets, we find that the improvement on the large datasets is not as significant as on the small dataset.
We think this phenomenon is quite reasonable due to the differences between the two datasets.
The bugs in the large dataset are usually simpler than those in the small dataset and can be fixed with single-line operations (discussed in Section~\ref{sec:rq1}).
As a result, {\toolname} has already achieved very high results on the large dataset (e.g., exceeding 99\% in all five metrics), which leads to the difficulty in obtaining even greater improvements.
Compared to the large dataset, we believe that the small dataset is able better to reflect the true capability of our approach and baselines, as the small dataset is constructed from Defects4J, which contains a variety type of real-world bugs and is the most widely used benchmark in the field of program repair and patch correctness assessment.
From Table~\ref{tab:training}, we observe a significant improvement in the dataset, which is more valuable to demonstrate the significance of our three training components.
}

\begin{table*}[htbp]
  \centering
  \renewcommand\arraystretch{1.1}
  \caption{\revise{Effectiveness of APPT with different training choices}}
    \begin{tabular}{c|c|c|c|c|c|c}
    \toprule
    \revise{Dataset} & \revise{Component} & \revise{Accuracy} & \revise{Precision} & \revise{Recall} & \revise{F1-score} & \revise{AUC} \\
    \midrule
    \multirow{4}[2]{*}{\revise{Small}} & \revise{{\toolname}$_{\textit{pre-training}^-}$}  & \revise{72.46\%(↑7.26\%)} & \revise{69.28\%(↑11.56\%)} & \revise{89.15\%(↑-5.98\%)} & \revise{77.97\%(↑3.8\%)} & \revise{82.24\%(↑0.31\%)} \\
          & \revise{{\toolname}$_{\textit{fine-tuning}^-}$}  & \revise{69.63\%(↑10.09\%)} & \revise{67.67\%(↑13.16\%)} & \revise{81.35\%(↑1.82\%)} & \revise{71.54\%(↑10.22\%)} & \revise{71.77\%(↑10.78\%)} \\
          & \revise{{\toolname}$_{\textit{LSTM}^-}$} & \revise{76.84\%(↑2.88\%)} & \revise{79.2\%(↑1.64\%)} & \revise{76.15\%(↑7.02\%)} & \revise{77.65\%(↑4.11\%)} & \revise{80.65\%(↑1.90\%)} \\
          & \revise{APPT}  & \revise{79.72\%} & \revise{80.84\%} & \revise{83.17\%} & \revise{81.76\%} & \revise{82.55\%} \\
    \midrule
    \multirow{4}[2]{*}{\revise{Large}} & \revise{{\toolname}$_{\textit{pre-training}^-}$} & \revise{98.\%(↑1.14\%)} & \revise{98.62\%(↑0.47\%)} & \revise{97.24\%(↑1.89\%)} & \revise{97.91\%(↑1.19\%)} & \revise{99.46\%(↑0.40\%)} \\
          & \revise{{\toolname}$_{\textit{fine-tuning}^-}$} & \revise{98.29\%(↑0.85\%)} & \revise{98.72\%(↑0.37\%)} & \revise{97.74\%(↑1.39\%)} & \revise{98.23\%(↑.88\%)} & \revise{99.74\%(↑0.12\%)} \\
          & \revise{{\toolname}$_{\textit{LSTM}^-}$} & \revise{86.66\%(↑12.48\%)} & \revise{85.94\%(↑13.15\%)} & \revise{86.7\%(↑12.42\%)} & \revise{86.31\%(↑12.8\%)} & \revise{93.35\%(↑6.51\%)} \\
          & \revise{APPT}  & \revise{99.13\%} & \revise{99.09\%} & \revise{99.13\%} & \revise{99.11\%} & \revise{99.86\%} \\
    \bottomrule
    \end{tabular}%
  \label{tab:training}%
\end{table*}%

\subsubsection{The Impact of The Vector Concatenation Choice}

\begin{table*}[htbp]
  \centering
    \caption{\newrevise{Effectiveness of {\toolname} with different concatenation choices}}
    \begin{tabular}{c|c|ccccc}
    \toprule
    Dataset & Concatenation & Accuracy & Precision & Recall & F1-score & AUC \\
    \midrule
    \multirow{6}[1]{*}{Small} & {\toolname}$_{\textit{addition}}$ & 69.83\% & 70.24\% & 80.12\% & 73.83\% & 75.44\% \\
          & {\toolname}$_{\textit{subtraction}}$  & 71.38\% & 72.42\% & 77.27\% & 74.72\% & 75.59\% \\
          & {\toolname}$_{\textit{product}}$  & 63.27\% & 62.37\% & \textbf{96.32\%} & 74.81\% & 66.46\% \\
          & \revise{{\toolname}$_{\textit{CACHE}}$}  & \revise{78.36\%} & \revise{79.33\%} & \revise{81.95\%} & \revise{80.60\%} & \revise{81.57\%} \\
          & \revise{{\toolname}$_{\textit{Tian et al.}}$} & \revise{78.19\%} & \revise{78.12\%} & \revise{84.09\%} & \revise{80.86\%} & \revise{81.67\%} \\
          & {\toolname}$_{\textit{concat}}$  & \textbf{79.72\%} & \textbf{80.84\%} & 83.17\% & \textbf{81.76\%} & \textbf{82.55\%} \\
    \midrule
    \multirow{6}[1]{*}{Large} & {\toolname}$_{\textit{addition}}$ & 98.96\% & 98.80\% & 99.07\% & 98.93\% & 99.81\% \\
          & {\toolname}$_{\textit{subtraction}}$  & 97.31\% & 99.14\% & 95.29\% & 97.17\% & 99.46\% \\
          & {\toolname}$_{\textit{product}}$  & 98.82\% & 98.88\% & 98.69\% & 98.78\% & 99.78\% \\
          & \revise{{\toolname}$_{\textit{CACHE}}$}  & \revise{99.07\%} & \revise{\textbf{99.19\%}} & \revise{99.11\%} & \revise{99.05\%} & \revise{98.84\%} \\
          & \revise{{\toolname}$_{\textit{Tian et al.}}$}  & \revise{99.06\%} & \revise{99.03\%} & \revise{\textbf{99.15\%}} & \revise{99.04\%} & \revise{98.85\%} \\
          & {\toolname}$_{\textit{concat}}$  & \textbf{99.13\%} & 99.09\% & \text{99.13\%} & \textbf{99.11\%} & \textbf{99.86\%} \\
    \bottomrule
    \end{tabular}%
  \label{tab:concatenation}%
\end{table*}%

\revise{
\emph{\textbf{Experimental Design.}}
In the vector integration process, {\toolname} directly concatenates the buggy method vector and patched method vector.
We formulate this subRQ to investigate the impact of different integration methods, such as concatenate additional, subtraction, and product operation \revise{(detailed in Section \ref{sec:classification})}.
We also consider the vector integration methods used in Tian et al. \cite{tian2020evaluating} and CACHE \cite{lin2021context}.
}

\emph{\textbf{Results.}}
Table \ref{tab:concatenation} presents the prediction results under different concatenation choices.
The first column lists the two datasets.
The second column lists the six concatenation choices, i.e., addition, subtraction, product, and the one used in CACHE and Tian et al. and APPT.
The remaining columns list the detailed values of accuracy, precision, recall, F1-score and AUC metrics.

On the small dataset,  although conceptually simple, {\toolname}$_{concat}$ can obtain 79.72\%, 80.84\%, 83.17\%, 81.76\%, and 82.55\% for accuracy, precision, recall, F1-score and AUC metrics, four of which are highest among all investigated concatenation methods.
{\toolname}$_{product}$ has the highest recall score (96.32\%), while it performs worse than {\toolname}$_{concat}$ by 16.45\%, 18.47\%, 6.95\% and 16.09\% for the other four metrics.
{\toolname}$_{addition}$ and {\toolname}$_{subtraction}$ perform the addition and subtraction operation for buggy and patched vectors, and have similar performance for all metrics.
Meanwhile, mixed methods (i.e., {\toolname}$_{\textit{CACHE}}$ and {\toolname}$_{\textit{Tian et al.}}$) that apply different comparison functions to represent the changed embedding vector can achieve comparable results with {\toolname}$_{concat}$, which is also consistent with the existing study results \cite{hoang2020cc2vec, tian2020evaluating}.
On the large dataset, {\toolname}$_{concat}$ achieves the best performance in accuracy, F1-score and AUC metrics, while {\toolname}$_{\textit{CACHE}}$ and {\toolname}$_{\textit{Tian et al.}}$) perform best in precision and recall respectively.
The difference in performance is similar as the methods have relatively high metric values.
For example,  all metric values are higher than 99\% for {\toolname}$_{concat}$ and {\toolname}$_{\textit{CACHE}}$.

\subsubsection{The Impact of Pre-trained Model Choice}

\begin{table*}[htbp]
\centering
\caption{\newrevise{Effectiveness of {\toolname} with different pre-trained models}}
\label{tab:rq3_models}
\begin{tabular}{c|l|ccccc}
    \toprule
    Dataset & Model & Accuracy & Precision & Recall & F1-score & AUC \\
    \midrule
    \multirow{3}[2]{*}{Small} & APPT$_{bert}$ & 79.72\%($\uparrow$ 4.32\%) & 80.84\%($\uparrow$ 1.34\%) & 83.17\%($\uparrow$ 6.67\%) & 81.76\%($\uparrow$ 3.76\%) & 82.55\%($\uparrow$ 2.25\%) \\
          & APPT$_{codebert}$ & 81.49\%($\uparrow$ 6.09\%) & 82.10\%($\uparrow$ 2.60\%) & 84.73\%($\uparrow$ 8.23\%) & 83.35\%($\uparrow$ 5.35\%) & 85.32\%($\uparrow$ 5.02\%) \\
          & APPT$_{graphcodebert}$ & 81.83\%($\uparrow$ 6.43\%) & 83.68\%($\uparrow$ 4.18\%) & 83.63\%($\uparrow$ 7.13\%) & 83.47\%($\uparrow$ 5.47\%) & 85.79\%($\uparrow$ 5.49\%) \\
    \midrule
    \multirow{3}[2]{*}{Large} & APPT$_{bert}$ & 99.13\%($\uparrow$ 0.53\%) & 99.09\%($\uparrow$ 0.19\%) & 99.13\%($\uparrow$ 0.93\%) & 99.11\%($\uparrow$ 0.51\%) & 99.86\%($\uparrow$ 0.96\%) \\
          & APPT$_{codebert}$ & 99.57\%($\uparrow$ 0.97\%) & 99.71\%($\uparrow$ 0.81\%) & 99.40\%($\uparrow$ 1.20\%) & 99.55\%($\uparrow$ 0.95\%) & 99.89\%($\uparrow$ 0.99\%) \\
          & APPT$_{graphcodebert}$ & 99.61\%($\uparrow$ 1.01\%) & 99.61\%($\uparrow$ 0.71\%) & 99.59\%($\uparrow$ 1.39\%) & 99.60\%($\uparrow$ 1.00\%) & 99.90\%($\uparrow$ 1.00\%) \\
    \bottomrule

\multicolumn{7}{l}{\small $\uparrow$ denotes performance improvement against state-of-the-art technique CACHE.}\\

\end{tabular}
\end{table*}

\revise{
\emph{\textbf{Experimental Design.}}
Recently, following the BERT model architecture, researchers use some code-related pre-trained tasks to capture the semantic connection between natural language and programming language, so as to further adapt these pre-training models for programming language.
Thus, we formulate this subRQ to investigate the impact of different pre-trained models by replacing the BERT with two advanced models pre-trained with the programming language, i.e., CodeBERT \cite{feng2020codebert} and GraphCodeBERT \cite{guo2020graphcodebert}.
}

\emph{\textbf{Results.}}
Table \ref{tab:rq3_models} demonstrates the predicted performance of three pre-trained models.
The first column lists the two datasets.
The second column lists the three models, i.e., BERT, CodeBERT, and GraphCodeBERT.
The remaining columns list the detailed values of accuracy, precision, recall, F1-score, and AUC metrics.

Generally speaking, all of the adopted models achieve a higher performance than the state-of-the-art technique CACHE on all metrics.
For example, on the small dataset, BERT, CodeBERT, and GraphCodeBERT reach 81.76\%, 83.35\%, and 83.47\% with respect to the F1-score, which is 3.76\%, 5.35\%, and 5.47\% higher than CACHE, respectively.
A similar improvement can also be observed on the large dataset.
This demonstrates the model choice may not impact the performance dramatically, and pre-trained models can consistently achieve state-of-the-art performance.

Specifically, to compare the performance of different pre-trained models, we can observe that both CodeBERT and GraphCodeBert achieve a better value for all metrics on the small dataset.
This superior performance also generalizes to large datasets, where CodeBERT and GraphCodeBert have better or competitive (e.g., AUC) performance on the metrics.
One possible explanation for this is that BERT is designed for natural language processing tasks, while CodeBERT and GraphCodeBERT regard a source code as a sequence of tokens or graph representation and then pre-train models on source code to support code-related tasks.
This indicates that although pre-trained models in NLP can achieve state-of-the-art performance for assessing patch correctness, the adoption of pre-trained models targeting source code can further boost the improvement.

\finding{2}{The performance under different choices demonstrates that:
\revise{
(1) all training components (e.g., fine-tuning and LSTM stack) positively contribute to {\toolname};}
\delete{
(2) the beginning code tokens can represent the buggy and patched code snippets well for the pre-trained model;}
(2) the concat of the buggy and patched vectors is better than other methods to distinguish the changed code snippets;
(3) advanced pre-trained models can provide a stable even better performance.
}

\subsection{\revise{RQ3: Cross-Project Prediction}}
\label{sec:cross}
\revise{
\emph{\textbf{Experimental Design.}}
\newrevise{We have demonstrated that {\toolname} achieves optimal performance in a cross-validation setting, which is the common practice in the APCA community.
It is possible that the patches generated by different APR tools for the same bug are split into the training set and testing set. 
However, in a real-world scenario, it would be impossible to use the label from one patch to predict the other for the same bug.}
In this section, to further provide insights about {\toolname}, we investigate the ability of {\toolname} to predict patch correctness in a cross-project setting.
In particular, we test all projects in Defects4J separately by creating training and testing sets per project.
For example, when we test all patches from the project Chart, we use the patches from other projects as the training set to ensure
the testing patches are new and unseen.
Due to the space limitation, we choose CACHE \cite{lin2021context} and Tian et al. \cite{tian2020evaluating} as the baselines, the former represents state-of-the-art and the latter is the most related work to our APPT.
}
\begin{table*}[htbp]
  \centering
  \caption{\revise{Effectiveness of {\toolname} in a cross-project setting}}
    \begin{tabular}{c|c|ccccc}
    \toprule
    \revise{Project} & \revise{Approach} & \revise{Accuracy} & \revise{Precision} & \revise{Recall} & \revise{F1-score} & \revise{AUC} \\
    \midrule
    \multirow{3}[2]{*}{\revise{Chart}} & \revise{Tian et al.} & \revise{58.90\%} & \revise{39.66\%} & \revise{41.82\%} & \revise{40.71\%} & \revise{55.64\%} \\
          & \revise{CACHE} & \revise{72.44\%} & \revise{\textbf{71.59\%}} & \revise{63.00\%} & \revise{67.02\%} & \revise{\textbf{78.94\%}} \\
          & \revise{APPA} & \revise{\textbf{73.49\%}} & \revise{71.01\%} & \revise{\textbf{76.44\%}} & \revise{\textbf{73.61\%}} & \revise{74.99\%} \\
    \midrule
    \multirow{3}[2]{*}{\revise{Closure}} & \revise{Tian et al.} & \revise{60.20\%} & \revise{61.64\%} & \revise{62.03\%} & \revise{61.83\%} & \revise{62.52\%} \\
          & \revise{CACHE} & \revise{64.00\%} & \revise{61.45\%} & \revise{51.00\%} & \revise{55.74\%} & \revise{\textbf{68.95\%}} \\
          & \revise{APPA} & \revise{\textbf{66.87\%}} & \revise{\textbf{69.29\%}} & \revise{\textbf{89.81\%}} & \revise{\textbf{78.23\%}} & \revise{63.18\%} \\
    \midrule
    \multirow{3}[2]{*}{\revise{Lang}} & \revise{Tian et al.} & \revise{63.37\%} & \revise{65.34\%} & \revise{62.78\%} & \revise{64.03\%} & \revise{68.16\%} \\
          & \revise{CACHE} & \revise{68.00\%} & \revise{66.28\%} & \revise{57.00\%} & \revise{61.29\%} & \revise{\textbf{75.76\%}} \\
          & \revise{APPA} & \revise{\textbf{72.98\%}} & \revise{\textbf{71.62\%}} & \revise{\textbf{71.00\%}} & \revise{\textbf{71.31\%}} & \revise{73.88\%} \\
    \midrule
    \multirow{3}[2]{*}{\revise{Math}} & \revise{Tian et al.} & \revise{58.22\%} & \revise{47.06\%} & \revise{49.72\%} & \revise{48.35\%} & \revise{57.03\%} \\
          & \revise{CACHE} & \revise{\textbf{69.78\%}} & \revise{69.62\%} & \revise{55.56\%} & \revise{61.80\%} & \revise{\textbf{76.33\%}} \\
          & \revise{APPA} & \revise{69.11\%} & \revise{\textbf{70.55\%}} & \revise{\textbf{84.25\%}} & \revise{\textbf{76.79\%}} & \revise{63.91\%} \\
    \midrule
    \multirow{3}[2]{*}{\revise{Time}} & \revise{Tian et al.} & \revise{71.11\%} & \revise{86.96\%} & \revise{66.67\%} & \revise{75.47\%} & \revise{75.33\%} \\
          & \revise{CACHE} & \revise{70.83\%} & \revise{\textbf{71.88\%}} & \revise{65.71\%} & \revise{68.66\%} & \revise{75.33\%} \\
          & \revise{APPA} & \revise{\textbf{80.00\%}} & \revise{71.43\%} & \revise{\textbf{66.67\%}} & \revise{\textbf{68.97\%}} & \revise{\textbf{83.11\%}} \\
    \midrule
    \multirow{3}[2]{*}{\revise{All}} & \revise{Tian et al.} & \revise{62.36\%} & \revise{60.13\%} & \revise{56.60\%} & \revise{58.31\%} & \revise{63.73\%} \\
          & \revise{CACHE} & \revise{69.01\%} & \revise{68.16\%} & \revise{58.45\%} & \revise{62.94\%} & \revise{\textbf{75.06\%}} \\
          & \revise{APPA} & \revise{\textbf{72.49\%}} & \revise{\textbf{70.78\%}} & \revise{\textbf{77.63\%}} & \revise{\textbf{74.05\%}} & \revise{71.81\%} \\
    \bottomrule
    \end{tabular}%
  \label{tab:cross-project}%
\end{table*}%

\revise{
\emph{\textbf{Results.}}
Table \ref{tab:cross-project} presents the effectiveness {\toolname} in a cross-project prediction scenario.
The first column lists the name of the project. 
The second column lists the investigated APCA techniques. 
The third to seventh columns list the results of accuracy, precision, recall, F1-score and AUC.
We also present the results by considering all five projects in the last row.
The best results per project are shown in bold.
}

\revise{
From the table, we can observe that {\toolname} still substantially outperforms compared techniques by achieving 72.49\% accuracy and 70.78\% precision, i.e., 3.48\% and 2.62\% more than CACHE.
Moreover, Recall and F1-score are consistently improved at least by 19.18\%	 and 11.11\% compared to CACHE. 
In addition, we can observe that compared to within-project prediction (i.e., RQ1), all techniques perform worse in the cross-project prediction scenario. 
For example, CACHE can predict 75.4\% patches correctly while only 69.01\% patches across different projects.
As for {\toolname}, it can predict 79.0\% patches within projects while 72.49\% patches across different projects. 
The observation is as expected, since in the within-project prediction scenario, testing data and training data may be from the same project, which tend to share similar features; whereas the cross-project prediction can be more challenging since characteristics between projects can be very different.
Even though, we can observe that compared to other techniques, {\toolname} exhibits the smallest effectiveness drop between within-project and cross-project prediction.
In summary, our results demonstrate that even when trained in the cross-project prediction scenario, {\toolname} still consistently outperforms state-of-the-art APCA techniques on hundreds of extra bugs.
}

\revise{
\finding{3}{The performance under a cross-project scenario demonstrates that:
(1) all investigated techniques show some decline in prediction performance compared with a cross-validation setting;
(2) {\toolname} still achieves optimal performance among \newdelete{all metrics}\newrevise{four of five investigated metrics} when predicting patch correctness from other projects.
}}

\section{Discussion}
\subsection{Threats to Validity}

To facilitate the replication and verification of our experiments, we have made the relevant materials (including source code, trained models, and patch data) available.
Despite that, our study still faces some threats to validity, listed as follows.

\newdelete{The first threat to validity lies in the patch benchmark.
We focus on the Defects4J database with reproducible real faults and collect 1,183 patches generated by existing APR tools.
However, the patch benchmark may not consider all available APR tools.
To address this, following the latest work \cite{lin2021context}, we include the 22 APR tools covering four categories.
It should be worth noting that although the learning-based category contains only SequenceR, it contains 73 patches, which is the largest number for a single APR tool \cite{lin2021context}.
We also mitigate the potential bias by using multiple evaluation metrics to exhaustively assess the APCA techniques.
Further, we adopt another large benchmark containing 49,694 real-world patches to evaluate the  generalization ability of the studied techniques.
Overall, to the best of our knowledge, the used patch benchmarks are the largest set explored in the literature on patch correctness assessment.}

\newrevise{The first threat to validity lies in the patch benchmark.
We focus on the Defects4J database with reproducible real faults and collect 1,183 patches generated by existing APR tools.
However, the patch benchmark may not consider all available APR tools.
To address this, following the latest work \cite{lin2021context}, we include the 22 APR tools covering four repair categories.
We also adopt the large benchmark containing 49,694 real-world patches and multiple evaluation metrics to evaluate the generalization ability of the studied techniques.
Another potential issue is that the patch benchmark may contain mislabelled patches.
In fact, these patches are manually labeled by the authors of the corresponding repair approaches, and then released for continuous review and correction by the APR community.
There is also a re-checking process when they are collected for APCA research.
Thus, we are confident in the reliability of these collected patches in our work, which have been employed by most of previous patch correctness studies~\cite{tian2020evaluating,tian2022change,tian2022predicting}.
Overall, to the best of our knowledge, the used patch benchmarks are the largest set explored in the literature on patch correctness assessment.}

The second threat to validity is that the performance of {\toolname} may not generalize to other pre-trained models.
We select BERT in our experiment due to its powerful performance in recent code-related works.
However, it is unclear whether the conclusions in our experiment (discussed in Section \ref{sec:re&an}) can be maintained when using other pre-trained models.
We have mitigated the potential threat by using CodeBERT and GraphCodeBERT to demonstrate the performance of {\toolname} under different pre-trained models.
The investigated pre-trained models include both code-related ones (e.g., CodeBERT) and natural language-specific ones (e.g., BERT).
We also rely on two diverse patch benchmarks to ensure the generality of the experimental conclusions.

The last threat to validity is the implementation of the baselines.
In our work, we compare {\toolname} against a wide range of APCA techniques with different categories.
Implementing these baselines may introduce a potential threat to the internal validity.
To mitigate this threat,  following the recent work \cite{lin2021context}, we conduct the experiment under the same setting and reuse the released results from the original work \cite{lin2021context, tian2020evaluating,wang2020automated}.
Further, we carefully check the reused results and publicly release all our materials for further verification.

\subsection{Comparison with BATS}

\begin{table*}[htbp]
  \centering
  \caption{\newrevise{Comparison with a state-of-the-art learning-based APCA technique BATS}}
    \begin{tabular}{c|c|cccccc}
    \toprule
    APCA  & Threshold & \#Patch & Accuracy & Precision & +Recall & \revise{-Recall} & F1-score \\
    \midrule
    BATS (0) & 0     & 1278  & 52.50\% & 48.81\% & 62.82\% & \revise{43.69\%} & 54.94\% \\
    BATS (0.8) & 0.8   & 114   & 67.54\% & 63.16\% & 84.21\% & \revise{50.88\%} & 72.18\% \\
    \midrule
    \multicolumn{2}{c|}{\toolname} & 1278  & \textbf{85.05\%} & \textbf{83.39\%} & \textbf{84.38\%} & \textbf{\revise{85.67\%}} & \textbf{83.88\%} \\
    \bottomrule
    \end{tabular}%
  \label{tab:bats}
\end{table*}%

In our work, following some recent APCA work \cite{tian2020evaluating, wang2020automated}, \delete{30 }\revise{17} related APCA techniques with different categories (i.e., \delete{16 }\revise{2} representation learning-based ones, 9 dynamic-based ones, 4 static-based ones and 2 learning-based ones) are compared in our experiment (discussed in Section \ref{sec:re&an}).
To the best of our knowledge, the selected baselines are the largest set on patch correctness prediction in the literature.
However, there may exist other possible techniques that could have been used.
For example, the recent BATS \cite{tian2022predicting} predicts patch correctness based on the similarity of failing test cases, which can be complementary to the state-of-the-art APCA techniques.
We do not include BATS in our experiment (discussed in Section \ref{sec:re&an}) because it requires historical test cases as the search space for searching similar cases, which are not available in our dataset.

We then perform an additional evaluation by assessing {\toolname} on the dataset provided in BATS \revise{by the standard 5-fold cross-validation}.
However, BATS fails to assess some plausible patches as it considers only historical test cases with the similarity which are higher than a threshold.
For example, BATS with 0.8 threshold value is able to predict only 8.9\% (114/1278) of the plausible patches.
Thus, we compare {\toolname} against BATS with 0.0 threshold value, which can perform prediction for all patches.
We also compare {\toolname} against BATS with 0.8 threshold value, as it achieves the best recall, F1-score and AUC performance among all threshold values.
The results are presented in Table \ref{tab:bats}.
The \delete{first column lists }\revise{first and second columns list} {\toolname} and BATS (with 0.0 and 0.8 threshold values, respectively).
\newrevise{The third column lists the number of predicted patches under the same testing set.
It is worth noting that BATS(0) and APPT can predict all patches (i.e., 1278 patches), while BATS(0.8) may fail to predict some plausible patches if no test case satisfies the threshold value.}
Each cell is represented as $x(y)$, where $x$ is the number of patches predicted by {\toolname} and BATS and $y$ is the total number of patches in the dataset.
The remaining columns list the detailed performance under the metrics.
\revise{We also present the performance of +Recall and -Recall as BATS focuses on identifying both correct patches and overfitting patches.}

From Table \ref{tab:bats}, we can find {\toolname} achieves 83.39\%$\sim$85.05\%, improving the metrics by 21.56\%$\sim$34.58\% when compared with BATS (threshold is set to 0.0).
When the threshold of BATS is set to 0.8, {\toolname} can still improve the metrics by 12.40\% on average while predicting 91.1\% more plausible patches.
\revise{
Besides, APPT is able to achieve 84.38\% for +Recall and 85.67\% for -Recall, indicating its ability to predict both overfitting and correct patches.}
Overall, the results demonstrate that {\toolname} performs better than BATS in terms of the number of predicted patches and the prediction metrics.

\subsection{\revise{Correct Patch Identification}}

\begin{table}[htbp]
  \centering
  \caption{\revise{Effectiveness of {\toolname} in identifying correct patches}}
    \begin{tabular}{c|ccccc}
    \toprule
    \revise{Approach} & \revise{-Accuracy} & \revise{-Precision} & \revise{-Recall} & \revise{-F1-score} & \revise{-AUC} \\
    \midrule
    \revise{Tian et al.} & \revise{64.84\%} & \revise{62.47\%} & \revise{55.70\%} & \revise{58.89\%} & \revise{55.64\%} \\
    \revise{CACHE} & \revise{76.33\%} & \revise{72.81\%} & \revise{76.07\%} & \revise{74.41\%} & \revise{74.99\%} \\
    \revise{APPA} & \revise{\textbf{79.12\%}} & \revise{\textbf{77.17\%}} & \revise{\textbf{76.45\%}} & \revise{\textbf{76.81\%}} & \revise{\textbf{78.94\%}} \\
    \bottomrule
    \end{tabular}%
  \label{tab:correct}%
\end{table}%

\revise{
In our work, following existing studies, we develop the pipeline of {\toolname} to filter the overfitting patches, so as to improve the precision of APR-generated patches.
The experimental results show {\toolname} achieves promising performance. 
To further evaluate {\toolname} in practice, we employ {\toolname} to identify the correct patches.
We employ the same evaluation metrics as the overfitting patch filtering scenario (discussed in Section \ref{sec:metric}).
It is worth noting that here the positive samples are correct patches and the negative samples are overfitting patches.
For example, the recall measures to what extent correct patches are identified, i.e., the percentage of correct patches that are indeed predicted as correct.
According to the definition of TP, FP, TN and FN, the Recall value is calculated by $-Recall = \frac{TN} {TN+FP}$.
Similar to Section \ref{sec:cross}, we select the state-of-the-art technique CACHE \cite{lin2021context} and the most related technique Tian et al. \cite{tian2020evaluating}.
\label{sec:cross}
}

\revise{
Table \ref{tab:correct} presents the comparison results.
We can find that {\toolname} still achieves better performance with respect to all metrics when identifying the correct patches.
For example, {\toolname} reaches 79.02\% accuracy, 77.17\% precision, 76.45\% recall, 76.81\% F1-score, and 78.94\% AUC,
which are 2.79\%, 4.36\%, 0.37\%, 2.40\%, and 3.95\% higher than the state-of-the-art CACHE.
The improvement against Tian et al. reaches 14.29\% $\sim$ 23.30\% on all metrics.
Such results further reflect the promising results of {\toolname} in the correct patch identification scenario.
}

\section{Implication and Guideline}

Based on the observations in our experiment, we can summarize the following essential practical guidelines for future patch correctness assessment studies.

\textbf{\delete{Simple }\revise{Sequence} features can work.}
\delete{
Our study demonstrates that {\toolname}, representing source code as a sequence of tokens, performs even better than the existing learning techniques (e.g., CACHE) considering complex code-aware characteristics (e.g., abstract syntax tree).
Also, the token sequences can already outperform manually-designed static features (e.g., the line number) and time-consuming dynamic features (e.g., code coverage) in this work.}
\revise{
Our study demonstrates that {\toolname}, representing source code as a sequence of tokens, performs even better than structure-based features considering complex code-aware characteristics (e.g., abstract syntax tree in CACHE) and manually-designed features (e.g., ODS).}
Such observations indicate that \delete{simple }\delete{sequence-based} features, such as code tokens, should not be just ignored and a systematic study to explore the impact of different code representations is needed in the future.
In fact, they should be considered and even integrated with different features (e.g., data flow graph and abstract syntax tree) to design more advanced patch correctness assessment techniques.

\textbf{The quality of the training dataset is important.}
We can find that {\toolname} achieves 91.5\% precision in Table \delete{\ref{tab:rq1_small_dataset}} \revise{\ref{tab:rq2_effectiveness}} \revise{(involving 902 patches)} while the precision is decreased by 10.8\% in Table \delete{\ref{tab:rq2_effectiveness}} \revise{\ref{tab:rq1_small_dataset}} \revise{(involving 1183 patches)}.
Similar performance can also be observed in Lin et al. \cite{lin2021context}.
The results show that more training data cannot always lead to better performance for patch correctness assessment.
It is crucial to automatically select the most informative training set that represents the whole patch benchmarks to optimize the prediction accuracy.
For example, it is interesting to explore how the number of patches is distributed across fix patterns and how to select balanced patches for each fix pattern.
Future work can also be conducted to investigate training data selection approaches targeting specific bug benchmarks under prediction or even specific bug types under prediction.

\textbf{Pre-trained model-based APCA techniques require more attention.}
Our results show that the BERT-based {\toolname} performs even better than the state-of-the-art APCA techniques.
Also, the CodeBERT-based and GraphCodeBERT-based {\toolname} can further enhance the prediction effectiveness.
Such observation motivates future researchers to investigate more advanced APCA techniques by employing different pre-trained models.
\delete{
For example, it is interesting to propose domain-specific pre-trained models by designing repair-related pre-training tasks.}
\revise{
It is interesting to conduct comprehensive and in-depth work to evaluate pre-trained models' capabilities (e.g., ChatGPT\footnote{ChatGPT. \url{https://openai.com/blog/chatgpt}, accessed March 2023}) for the APCA task.}
Meanwhile, thorough evaluations are recommended to explore how different features, such as bug types and fix patterns, influence the performance of pre-trained models in patch correctness prediction.

\section{Related Work}
\label{sec:rw}

\delete{In this paper, we adopt pre-trained language models to predict patch correctness generated by off-the-shelf automated program repair tools.}
\revise{In this paper, we adopt advanced pre-trained language models to address the important overfitting problem (i.e., the generated patches are overfitting to the given test suite) in automated program repair.}
Our work is related to automated program repair, patch correctness assessment and pre-trained models.
We have introduced the existing work about patch correctness assessment in Section \ref{sec:baselines}.
Thus, in this section, we focus on and discuss the existing work on automated program repair techniques (Section \ref{sec:apr}) and pre-trained models (Section \ref{sec:premodel}).

\subsection{Automated Program Repair}
\label{sec:apr}

\delete{
Over the past decade, researchers have proposed a variety of techniques to generate patches based on different hypotheses \cite{gazzola2017automatic, monperrus2020living}.
Following recent work \cite{zhang2022program, benton2021evaluating, liu2020efficiency}, we categorize them into four main categories: heuristic-based~\cite{le2012genprog, martinez2016astor,Yuan2018Arja}, constraint-based~\cite{durieux2016dynamoth,Xuan2016Nopol, mechtaev2016angelix}, template-based \cite{koyuncu2020fixminer,Liu2019Avatar,liu2019Tbar} and learning-based repair techniques~\cite{li2020dlfix,zhu2021syntax,lutellier2020coconut,yuan2022circle}.
}

\revise{{\toolname} investigates patches generated by existing APR techniques in the literature \cite{monperrus2020living}.
These techniques are from different categories, i.e., heuristic-based~\cite{le2012genprog, martinez2016astor}, constraint-based~\cite{durieux2016dynamoth,Xuan2016Nopol, mechtaev2016angelix}, template-based \cite{koyuncu2020fixminer,Liu2019Avatar} and learning-based repair techniques~\cite{li2020dlfix,zhu2021syntax}, summarized as follows.
}

\emph{$\bullet$ Heuristic-based repair techniques.}
These techniques usually use a heuristic algorithm to find a valid patch by iteratively exploring a search space of syntactic program modifications \cite{le2012genprog, martinez2016astor, Yuan2018Arja}.
Among them, GenProg \cite{le2012genprog} proposed in the early days has been considered a seminal work in this field, which uses genetic programming to search for correct repairs.
GenProg represents candidate repairs as sequences of edits to source code and evaluate them by the execution results of test cases.
Those candidates that pass more test cases are considered to have a higher fitness and are iteratively applied to produce new candidates based on mutation and crossover operations.
\revise{GenProg is reported to fix 55 of 115 considered bugs, but only 2 of them are fixed correctly, raising the overfitting issue in the APR community.}
The recent SimFix technique \cite{jiang2018shaping} utilizes code change operations from existing patches across different projects and similar code snippets within the buggy project to build two search spaces.
Then, the intersection of the above two search spaces is further used to search the final patch using basic heuristics.
\revise{SimFix is able to generate 34 correct patches along with 22 overfitting patches, leading to a precision of 60.7\%.}

\emph{$\bullet$ Constraint-based repair techniques.}
These techniques mainly focus on repairing conditional statements, which can repair more than half of the bugs repaired by existing APR approaches \cite{2018Martinez, durieux2016dynamoth, Xuan2016Nopol}.
In detail, these techniques transform the patch generation into a constraint-solving problem, and use a solver to obtain a feasible solution.
\delete{For example, Nopol \cite{Xuan2016Nopol} relies on an SMT solver to solve the condition synthesis problem after identifying potential locations of patches by angelic fix localization and collecting test execution traces of the program.}
\revise{For example, Nopol \cite{Xuan2016Nopol} relies on an SMT solver to solve the condition synthesis problem and generates 5 correct patches among 35 plausible patches for Defects4J \cite{martinez2017automatic}.}
\delete{Among them, ACS~\cite{2017Xiong} refining the ranking of ingredients for condition synthesis is considered one of the most advanced constraint-based repair techniques~\cite{liu2020efficiency}.}
\revise{
Furthermore, ACS~\cite{2017Xiong} generates patches that are highly likely to be correct by refining the ranking of ingredients for condition synthesis and successfully generates 18 correct patches over 23 generated patches with a precision of 78.3\%.
}

\emph{$\bullet$ Template-based repair techniques.}
These techniques generate patches by designing pre-defined fix patterns to mutate buggy code snippets with the retrieved donor code \cite{koyuncu2020fixminer, liu2019Tbar, Liu2019Avatar}.
For example, Liu et al. \cite{liu2019Tbar} revisit the repair performance of repair patterns (i.e., Tbar) using a systematic study that evaluates the effectiveness of a variety of fix patterns summarized from the literature.
\revise{
Tbar is able to generate 101 plausible patches for 101 bugs and 74 of them are considered to be fixed correctly (i.e., semantically identical to the human-written patch).
}
\delete{Among them, the recent PraPR technique \cite{ghanbari2019prapr} is able to generate plausible and correct patches for 148 and 43 real bugs, respectively, which is the largest number of bugs reported as fixed for Defects4J when published.}
\revise{PraPR \cite{ghanbari2019prapr} is able to generate plausible patches for 148 real-world Defects4J bugs using JVM bytecode mutation but only 43 bugs are correctly fixed.
}

\emph{$\bullet$ Learning-based repair techniques.}
These techniques attempt to fix bugs enhanced by machine learning techniques \cite{2019White, tufano2019empirical, gupta2017deepfix, lutellier2020coconut, li2020dlfix, jiang2021cure} and are getting increasing attention recently.
For example, Tufano et al. \cite{tufano2019empirical} extensively evaluate the ability of neural machine translation techniques to generate patches from bug-fixes commits in the wild.
\revise{
Chen et al. \cite{chen2019sequencer} propose a novel end-to-end approach (i.e., SEQUENCER) to program repair based on sequence-to-sequence learning.
SEQUENCER generates plausible patches for 19 bugs and 14 bugs are considered to be correctly fixed.}
Li et al.~\cite{li2020dlfix} adopt a tree-based RNN encoder-decoder model (i.e., DLFix) to learn code contexts and transformations from previous bug fixes.
Lutellier et al. \cite{lutellier2020coconut} propose a new context-aware NMT architecture (i.e., CoCoNut) that represents the buggy source code and its surrounding context separately, to automatically fix bugs in multiple programming languages.

\revise{We find that although APR is widely investigated, the correctness of APR-generated patches is still a long-standing challenge in the literature \cite{winter2022developers}.}
In our experiment, we select 22 representative APR tools (e.g., SimFix, ACS, and SEQUENCER) from the four categories, representing state-of-the-art techniques in the corresponding category.
Then we evaluate {\toolname} on the plausible patches (i.e., passing the original test cases) generated by these APR techniques.
 
\subsection{Pre-trained Model}
\label{sec:premodel}
Our approach is inspired by the application of pre-trained models in NLP and code-related tasks.
In this section, we first introduce the existing studies about pre-trained models in NLP (Section \ref{sec:nlp}) and SE (Section \ref{sec:se}).
We then discuss the application of pre-trained models to some code-related tasks in SE  (Section \ref{sec:other_se}).

\subsubsection{Pre-trained Model in NLP}
\label{sec:nlp}
Recent work has demonstrated substantial gains on many NLP tasks and benchmarks by pre-training on a large corpus of text followed by fine-tuning on a specific task.
For example, Devlin et al. \cite{devlin2018bert} propose a new language representation model BERT to pre-train deep bidirectional representations from the unlabeled text by jointly conditioning on both left and right contexts in all layers.
To explore the landscape of transfer learning techniques for NLP, Raffel et al. \cite{raffel2019t5} propose a text-to-text transfer transformer T5 by introducing a unified framework that converts all text-based language problems into a text-to-text format.
Brown et al. \cite{brown2020gpt} propose an autoregressive language model GPT-3 without any gradient updates or fine-tuning, with tasks and few-shot demonstrations specified purely via text interaction with the model.

In this work, we choose BERT to encode a given plausible patch into a fixed-length representation vector as the input of the deep learning classifier, due to the powerful performance of BERT in
previous work \cite{ciniselli2021empirical}.

\subsubsection{Pre-trained Model in SE}
\label{sec:se}
Inspired by the application of pre-trained models in NLP, many researchers apply the pre-trained model to code-related tasks.
Instead of designing new network architectures, SE researchers usually adopt existing architectures in NLP and design some code-aware pre-training tasks (e.g., code-AST prediction and bimodal dual generation) to learn representations of the source code.
Then the pre-trained models are further fine-tuned to some diversified code-related tasks such as code-code (clone detection, defect detection, cloze test, code completion, code refinement, and code-to-code translation), text-code (natural language code search, text-to-code generation), and code-text (code summarization) scenarios.

For example, Feng et al. \cite{feng2020codebert} present a bimodal pre-trained model (\textit{CodeBERT}) for natural language and programming languages by masked language modeling and replaced token detection to support code search and code documentation generation tasks.
Guo et al. \cite{guo2020graphcodebert} present the first pre-trained model (\textit{GraphCodeBERT}) that leverages code structure to learn code representation to improve code understanding tasks (i.e., code search, clone detection, code translation, and code refinement).
Guo et al. \cite{guo2022unixcoder} present UniXcoder, a unified cross-modal pre-trained model for programming language.
UniXcoder utilizes mask attention matrices with prefix adapters to control the behavior of the model and leverages cross-modal contents such as AST and code comment to enhance code representation.
In contrast to most studies pre-training a large-scale model from scratch costly, we attempt to boost patch correctness assessment on top of the existing pre-trained language model fine-tuning paradigm.

In this work, to further explore the generalization ability of {\toolname}, we select other BERT-like models (i.e., CodeBERT and GraphCodeBERT) as the encoder stack due to their powerful performance in the code-related tasks.

\subsubsection{Applications of Pre-trained Model in SE}
\label{sec:other_se}

In addition to the above-mentioned typical code-related tasks (e.g., automatic bug-fixing, injection of code mutants, generation of asserts in tests and code summarization in \cite{mastropaolo2021t5learning}), researchers have also applied pre-trained models to some other domains (e.g., code completion, and program repair~\cite{zhang2023critical,zhang2023gamma,zhang2023pre}) in SE.

For example, Cinisell et al. \cite{ciniselli2021empirical} evaluate the performance of the BERT model in the task of code completion at different granularity levels, including single tokens, one or multiple entire statements.
The results show that the model achieves promising results superior to state-of-the-art n-gram models, and the model learns better on some specific datasets (e.g., Android) when code abstraction is used.
Ciborowska et al. \cite{ciborowska2022fast} apply BERT to the bug localization problem with the goal of improved retrieval quality, especially on bug reports where straightforward textual similarity would not suffice.
Recently, Salza et al. \cite{salza2022effectiveness} investigate how transfer learning can be applied to code search by  pre-training and fine-tuning a BERT-based model on combinations of natural language and source code.
Mashhadi et al. \cite{mashhadi2021applying} propose a novel pre-trained model-based APR technique by fine-tuning CodeBERT on the ManySStuBs4J benchmark and find the approach generates fix codes for different types of bugs with comparable effectiveness and efficacy compared with state-of-the-art APR techniques.
\revise{
Recently, Tian et al.~\cite{tian2022change} treat the APCA problem as a question-answering task and propose a learning-based approach that exploits a deep NLP model to classify the relatedness of a bug report with a patch description based on a pre-trained BERT model.
}

Although there exist some SE tasks (e.g., code review and bug localization) benefitting from pre-trained models, in this work, we perform the first application of \revise{fine-tuning} pre-trained models to predict the generated patch correctness in automated program repair.

\section{Conclusion}
\label{sec:con}

In this work, we present {\toolname}, a novel pre-trained model-based automated patch correctness prediction technique by both pre-training and fine-tuning.
We first adopt the off-the-shelf pre-trained model as the encoder stack and LSTM stack to enhance the dependency relationships among the buggy and patched code snippets.
Then we build a deep learning classifier with two fully connected layers and a standard softmax function to predict whether the patch is overfitting or not.
We conduct experiments on both patch datasets and show that {\toolname} significantly outperforms state-of-the-art learning-based and traditional APCA techniques.
We further demonstrate that {\toolname} is generalizable to various pre-trained models.
Based on these observations, some implications and guidelines on improving the existing learning-based techniques (e.g., the usage of simple features and pre-trained models) are provided.
Overall, we highlight the direction of applying pre-trained models to predict patch correctness automatically.

\ifCLASSOPTIONcompsoc
  \section*{Acknowledgments}
\else
  \section*{Acknowledgment}
\fi
The authors would like to thank the anonymous reviewers for their insightful comments.
This work is supported partially by the National Key Research and Development Program of China (2021YFB1715600), the National Natural Science Foundation of China (61932012, 62141215, 62372228), and the China Scholarship Council (202306190117).

\ifCLASSOPTIONcaptionsoff
  \newpage
\fi

\bibliographystyle{IEEEtran}
\bibliography{reference}

\end{document}